\documentclass[fleqn,usenatbib]{mnras}

\usepackage{newtxtext,newtxmath}
\usepackage[T1]{fontenc}

\DeclareRobustCommand{\VAN}[3]{#2}
\let\VANthebibliography\thebibliography
\def\thebibliography{\DeclareRobustCommand{\VAN}[3]{##3}\VANthebibliography}

\usepackage{graphicx}	
\usepackage{amsmath}	
\usepackage{booktabs}
\usepackage{multirow}
\usepackage{subfigure} 
\usepackage{longtable}

\usepackage{adjustbox}
\usepackage{subcaption} 

\newcommand{\lya}{Ly$\alpha$}
\newcommand{\Oiii}{[O\,{\sc iii}]}
\newcommand{\lyb}{Ly$\beta$}

\title[Ly$\alpha$ Emission and Environment]{Ly$\alpha$ Emission from \Oiii\ Emitters Near Reionization:
The role of environment in galaxy Ly$\alpha$ detection}

\author[ Hashemi et al.]{
Seyedazim Hashemi,$^{1}$\thanks{E-mail: seyedazim.hashemi@email.ucr.edu}
George D. Becker,$^{1}$
Yongda Zhu,$^{2}$ 
Hui Hong$^{1}$
\\
$^{1}$Department of Physics \& Astronomy, University of California, Riverside, CA, 92521, USA\\
$^{2}$Steward Observatory, University of Arizona, 933 N Cherry Ave, Tucson, AZ 85721, USA\\
}

\date{Accepted XXX. Received YYY; in original form ZZZ}

\pubyear{\the\year{}}

\begin{document}
\label{firstpage}
\pagerange{\pageref{firstpage}--\pageref{lastpage}}
\maketitle

\begin{abstract}

Using galaxy \lya\ emission to probe reionization relies on establishing baseline expectations for its detectability in the absence of attenuation by neutral gas in the IGM.  Towards this end, the growing numbers of $z \sim 5$--6 star-forming galaxies spectroscopically selected by {\it JWST} provide an ideal sample for determining how \lya\ emission depends on galaxy properties and environment after reionization has largely completed.  In this study, we use Keck LRIS to measure the Ly$\alpha$ emission of 46 {\it JWST}-selected \Oiii-emitting galaxies over $5.3 \lesssim z \lesssim 6.2$ in the foreground of the ultra-luminous quasar J0100+2802.  Overall, we find that the fraction of galaxies detected in \lya\ emission is consistent with previous works; however, the fraction also varies with environment.  Most notably, we find an apparent deficit of \lya\ in the largest group in our sample, at $z \simeq 6.19$, which falls within the redshift range of the quasar's highly ionized proximity zone.  We speculate that the \lya\ emission from this group may be partly scattered by a foreground neutral island.  In contrast, we detect a high rate of \lya\ emission in two groups at $z \simeq 5.73$ and $z \simeq 5.78$.  These groups may be part of a structure that is extended along the line of sight, enhancing the transmission of Ly$\alpha$ emission.  While our sample size is limited, our results suggest that environment may play a significant role in the detectability of galaxy \lya\ emission even as late as $z \sim 6$.

\end{abstract}


\section{Introduction}

Over the past two decades a variety of observational tools have put strong constraints on the timing of hydrogen reionization.  Measurements of CMB electron scattering optical
depth imply a midpoint of reionization near $z \sim 7$--8 \citep[e.g.,][]{Planck2020}.  Damping wing absorption in the spectra of the most distant known quasars also support a large neutral fraction at $z \sim 7$ \cite[e.g.,][]{Mortlock2011, Greig2017, Banados2018, Davies2018, Greig2019,  Wang2020, Yang2020, Greig2022} that decreases toward $z \sim 6$ \citep{durovcikova2024,Greig2024}.  Recently, damping-wing absorption detected in galaxy spectra with {\it JWST} has also been used to probe the neutral fraction, supporting a fully neutral IGM at $z \gtrsim 9$ \citep{Umeda2024,Umeda2025b, Umeda2025a, Mason2025}.  Meanwhile, \lya\ forest absorption in quasar spectra indicate that while reionization is largely ($\gtrsim$80\%) complete by $z \sim 6$, islands of neutral gas may persist in the IGM as late as $z \sim 5.3$ (e.g., \citealt{Fan2006, McGreer2015, Becker2015b, Bosman2018, Eilers2018, Yang2020, Bosman2022, Zhu2022, Zhu2024, Spina2024, Becker2024}). In good agreement with quasar \lya\ forest studies, galaxies are also now being used to measure \lya\ forest absorption \citep{Meyer2025}.  

One of the most important tools for studying reionization is galaxy \lya\ emission \citep[e.g.,][]{Dijkstra2014}.
Strong Ly$\alpha$ emission is frequently observed in galaxies at redshift $z < 6$ (e.g., \citealt{Stark2011, Curtis-Lake2012, Cassata2015, DeBarros2017, Tang2024c}), but it becomes increasingly more rare at higher redshifts \citep[e.g.,][]{Schenker2012,Schenker2014,Treu2013,Pentericci2014,Pentericci2018, Tang2024b, Witstok2025}. The decline is commonly attributed to scattering by neutral hydrogen in the IGM.  Measurements of the galaxy \lya\ fraction can thus serve as a probe of reionization IGM at higher redshifts than are accessible with \lya\ forest absorption, which typically saturates at $z \gtrsim 6$.  The global galaxy \lya\ fraction can either be directly calculated by measuring the incidence of \lya\ emission among galaxies with independent redshift constraints \citep[e.g.,][]{Stark2011, Tang2024b}, or inferred by comparing the \lya\ luminosity function of \lya-selected galaxies (typically from narrow-band or blind spectroscopic surveys) to the UV luminosity function of continuum-selected galaxies \citep[e.g.,][]{Ouchi2010, Santos2016, Ota2017, Konno2018, Runnholm2025}.  Both methods indicate a decline in the fraction of galaxies with detectable \lya\ and a corresponding increase in the IGM neutral fraction that rises rapidly at $z > 6$ \citep[e.g.,][]{Schenker2012,Treu2013,Pentericci2014,Pentericci2018,Mason2018b, Mason2019, Whitler2020, Bolan2022, Goto2021, Tang2024b, Kageura2025}.  In addition to providing global constraints on the IGM, galaxy Ly$\alpha$ emission can be used to trace the growth of individual ionized bubbles (e.g., \citealt{Barkana2004, Furlanetto2004, Wyithe2005, Iliev2006, Dayal2018, Weinberger2018, Endsley2021,Trapp2023,Hayes2023,Lu2025}).

Multiple factors can impact the detectability of galaxy \lya. For example, galaxy gas dynamics may play a role.  If most of the Ly$\alpha$ photons from a galaxy are already significantly redshifted from their systematic redshift, its Ly$\alpha$ emission can pass through the IGM even if the galaxy is located in a small ionized region (e.g., \citealt{Erb2014, Willott2015, Mason2018, Hashimoto2019, Matthee2020, Endsley2022}). Furthermore, some galaxies may have an intense radiation field the facilitates the production and visibility of Ly$\alpha$ from small ionized bubbles (e.g., \citealt{Stark2017, Endsley2021, Simmonds2023, Tang2023, Roberts-Borsani2023}).  Recently, several studies have data from the {\it James Webb Space Telescope} ({\it JWST}) to investigate conditions favorable for detecting LAEs. In addition to being located within large ionized bubbles \citep[e.g.,][]{Witten2024, Witstok2024, Chen2024, Whitler2024}, \cite{Witten2024} suggested that galaxy mergers, which generate a significant number of Ly$\alpha$ photons, along with a ‘favorable’ line-of-sight—cleared of local neutral hydrogen in the host galaxy—are key factors in making Ly$\alpha$ emission detectable. The line-of-sight extension of galaxy groups has also been proposed as a potential factor in Ly$\alpha$ detection \citep{Chen2024}.

Increasing the robustness of galaxy \lya\ emission as a tracer of reionization requires establishing a robust baseline for the detectability of \lya\ in the absence of scattering by a neutral IGM.  In order to minimize the impact of galaxy evolution, this baseline should be set soon after reionization ends.  An ideal sample would include galaxies in a range of environments with secure spectroscopic redshifts measured independently from \lya.  Fortunately, several {\it JWST} programs are now providing such samples.  Surveys such as EIGER and ASPIRE are identifying large samples of \Oiii-selected galaxies  at $z \lesssim 6$ via NIRCam wide-field slitless spectroscopy (e.g., \citealt{Kashino2023, Matthee2023, Eilers2024, Wang2023}).  Because the surveys are conducted in the fields of background quasars, they also provide the opportunity to examine the connections between galaxies and the IGM properties traced by the quasar spectra \citep[e.g.,][]{Kashino2023,Jin2024,Kakiichi2025,Conaboy2025}.

In this paper we present a pilot study of the connection between \lya\ detectability and environment in \Oiii-selected galaxies at $z \lesssim 6$.  We use Keck LRIS to measure \lya\ the emission from a sample of galaxies identified in the EIGER survey by \cite{Kashino2023}.  The galaxies span a range of environment, from isolated to large groups.  Although our sample is relatively small, we find tentative evidence that environment may indeed play a role in \lya\ detectability even near the end of reionization. 

The organization of the remainder of the paper is as follows. In Section \ref{sec:lyameasurements}, we describe our sample of \Oiii-selected galaxies, LRIS observations  and data reduction.
In Section \ref{section:results}, we examine the   Ly$\alpha$ equivalent width ($W$) distribution and its variation among the groups in our sample.  We find that a large group at $z=6.19$ appears to show significantly less \lya\ than the remainder of the sample, and discuss possible explanations.  We also find that a two smaller groups at $z \simeq 5.7$--5.8 that are aligned along the line of sight exhibit a possible \lya\ enhancement.
Finally, we summarize our results and conclusions in Section \ref{section:summary}.  In this work, we assume a flat $\Lambda$CDM cosmology with $H_0=67.4~\mathrm{km~s^{-1}~Mpc^{-1}}$ and $\Omega_\mathrm{M}=0.315$ (\citealt{Planck2020}). All magnitudes are quoted in the AB system \cite{Oke1983}, and emission line equivalent widths are measured in the rest frame.

\section{\lya\ measurements}
\label{sec:lyameasurements}

\subsection{Target selection}
\label{subsec:oiiiselected}

Our targets are drawn from the sample of \Oiii-emitting galaxies in the field of SDSS J010013.02+280225.8 (hereafter J0100+2802) identified in the EIGER survey (JWST Program ID GO-1243, PI: Lilly).  EIGER uses JWST NIRCam imaging and wide-field slitless spectroscopy to identify line-emitting galaxies in the fields of six $z > 6$ quasars.  For \Oiii\ emission, the wavelength coverage spans $5.3 \lesssim z \lesssim 7 $.  The footprint is approximately $6.5 \times 3.4$ arcmin$^2$, or up to $\sim$10 cMpc/$h$ around the sight line.  \cite{Kashino2023} identify 117 \Oiii-emitting systems within this window towards J0100+2802. These galaxies span a wide range in UV ($-22.3\leq M_\mathrm{UV} \leq -17.7$) and \Oiii\ luminosity ($41.8\leq \log_{10}( L_{\mathrm{[O\ III]}}/ \text{erg s}^{-1}) \leq43.4$). The sample includes three prominent overdensities.   One of these is situated behind the quasar, at $z \simeq 6.78$.  The second includes J0100+2802 itself, while the third occurs at $z \simeq 6.19$, which coincides in redshift with the end of the quasar's proximity zone.  As argued by \cite{Matthee2023}, the narrow \Oiii\ line widths and extended morphologies suggest that the emission of these sources is typically powered by star formation.

\begin{figure*}
    \centering
    \includegraphics[width=.8\textwidth]{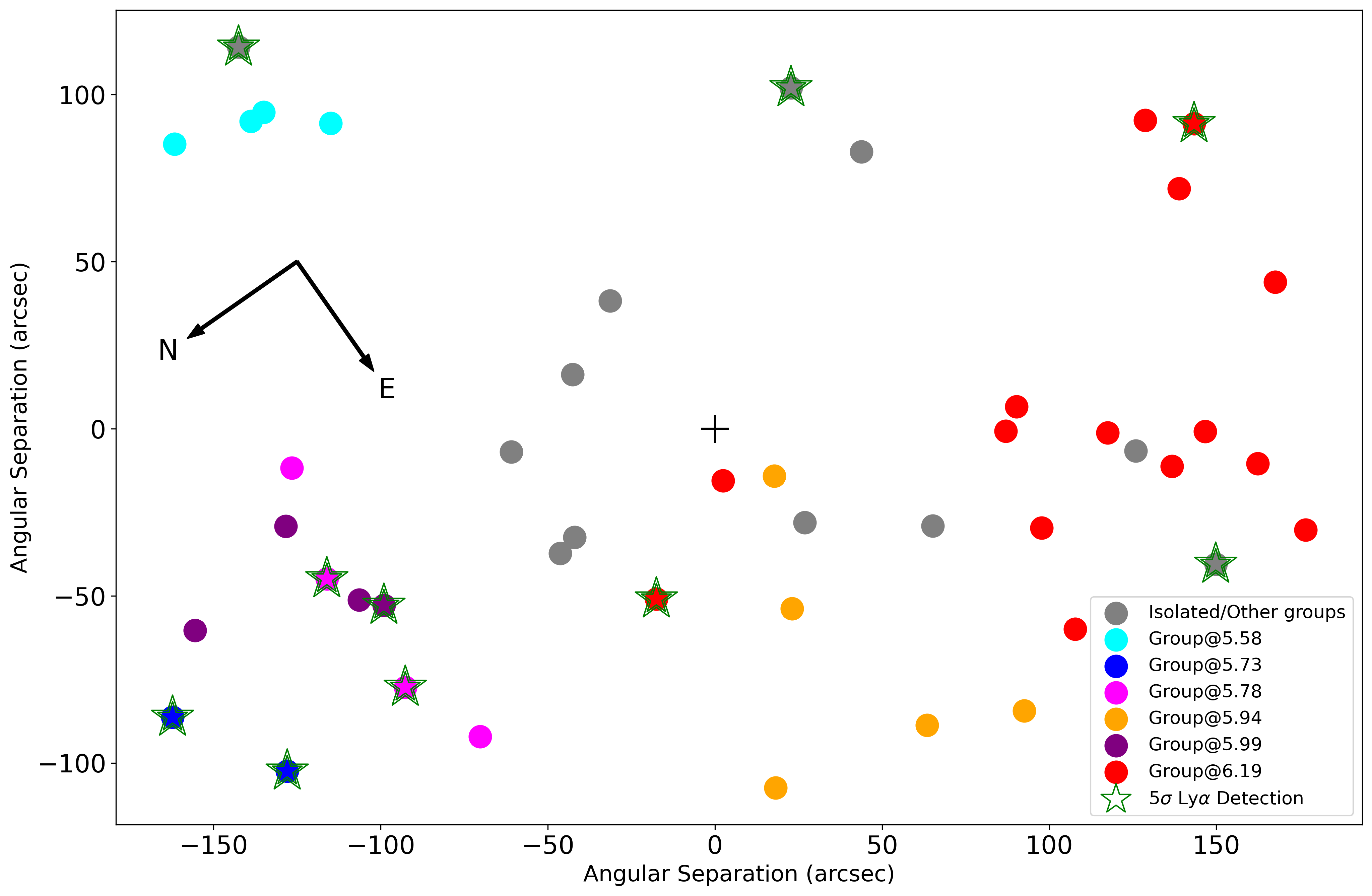}
    \caption{Spatial distribution of \Oiii\ emitters in the field of J0100+2802 targeted with Keck LRIS. Colors indicate membership in the groups listed in the legend. Gray points represent galaxies that are not part of any group in our sample; however, they may belong to a group in the parent \Oiii\ sample. The black cross at the center marks the position of the quasar. Green  stars denote sources where we detect Ly$\alpha$ emission at more than 5$\sigma$ confidence. Orientation is chosen to be similar to Figure 7 in \citealt{Kashino2023}.}
    \label{fig:sample_spatial_distribution}
\end{figure*}

\begin{figure*}
    \centering
        \includegraphics[width=.33\linewidth]{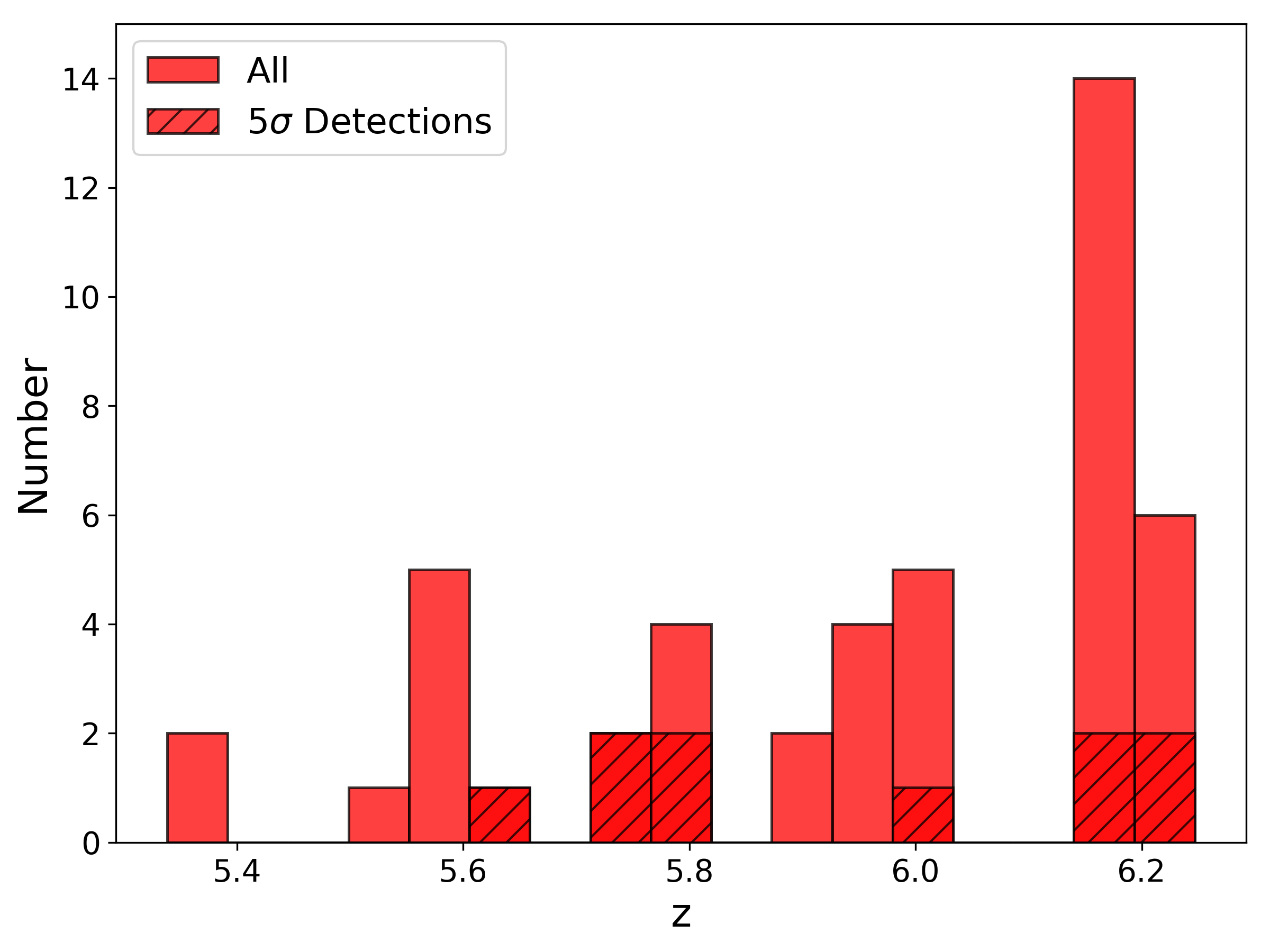}
        \includegraphics[width=.33\linewidth]{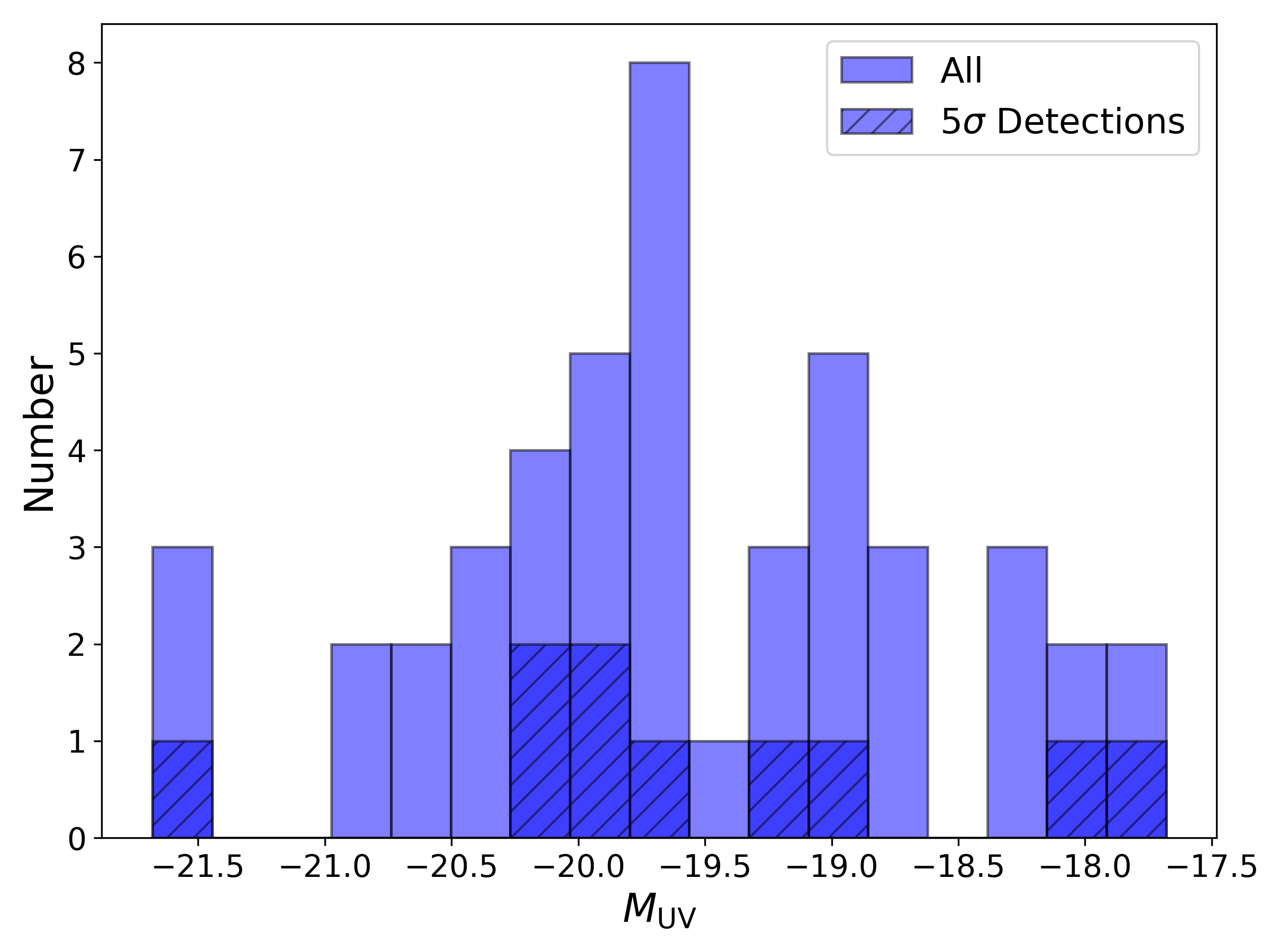}
        \includegraphics[width=.33\linewidth]{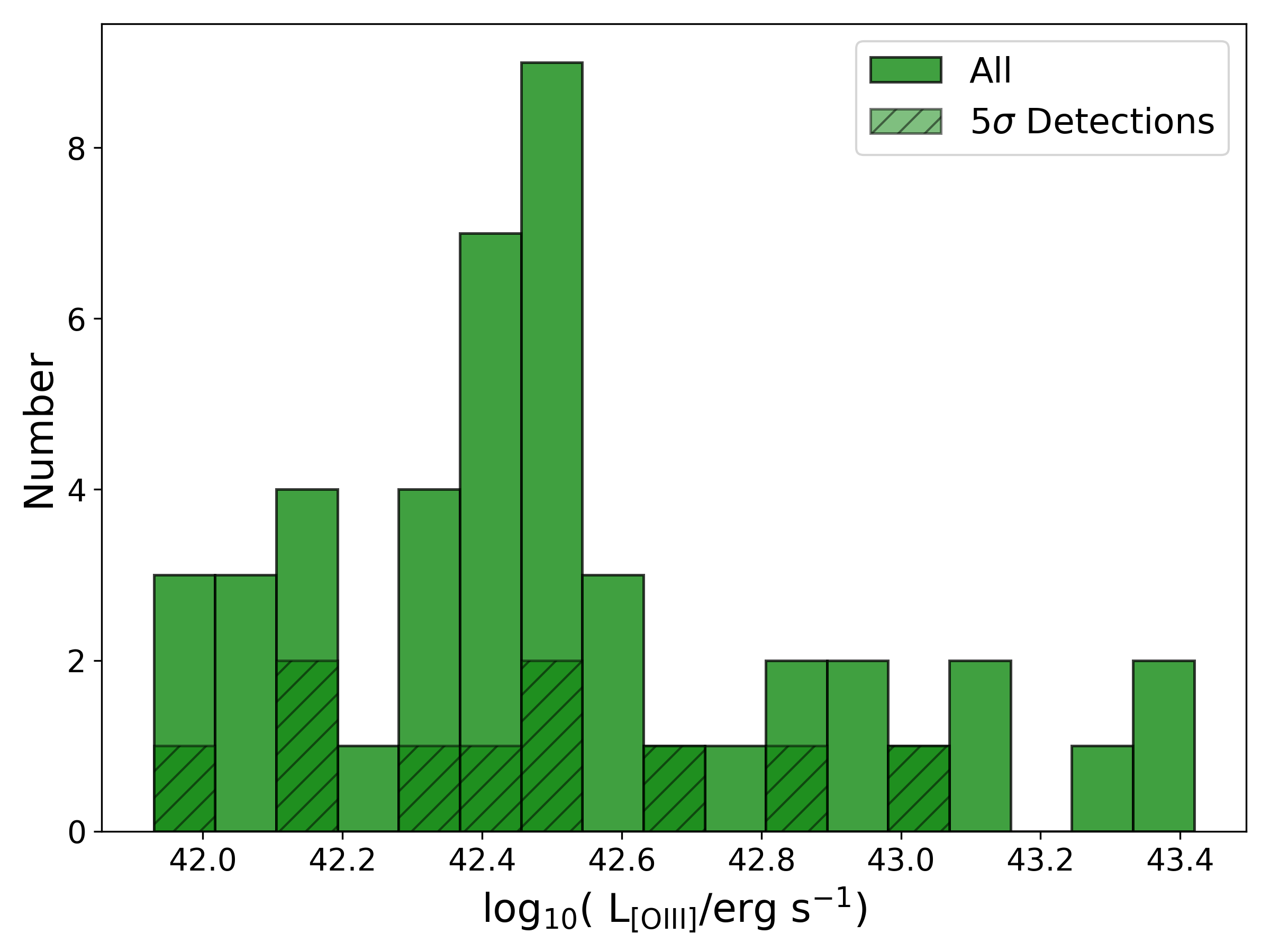}
    \caption{Distribution of redshift (left), $M_{\mathrm{UV}}$ (center) and \Oiii\ luminosity (right) of the galaxies in our sample.  Sources with \lya\ detected at $>$5$\sigma$ confidence are indicated with a hatched pattern.}
    \label{fig:sample_histogram}        
\end{figure*}

In this work we target 46 galaxies in the foreground of J0100+2802.  The galaxies were chosen to span a range of environments and redshifts and also are prioritized based on the confidence of their \Oiii\ detections. We also attempted to optimize the number of galaxies that could be observed using only two masks in order to increase the exposure time for each object. The spatial distribution of these galaxies relative to the central quasar is shown in Figure \ref{fig:sample_spatial_distribution}.  Distributions of redshift, \Oiii\ luminosity, and $M_\mathrm{UV}$ are presented in Figure \ref{fig:sample_histogram}. The $L_{\mathrm{[O\ III]}}$ and $M_\mathrm{UV}$ values of our targets span nearly the full range of the EIGER sample.  Based on their spatial distribution and similarity in redshift ($\Delta z < 0.02$), we identify six galaxy groups among our targets (see Figure~\ref{fig:sample_spatial_distribution}). Notably, our sample includes 15 objects from the large overdensity at $z \simeq 6.19$.   We return to this overdensity below.

\subsection{Keck LRIS data}
\label{subsec:kecklrisdata}

We observed our objects with Keck LRIS over two half-nights on November 12 (Night 1) and 13 (Night 2) in 2023. The observations were conducted using the 600/10000 grating on the red side with the 560 nm dichroic, and 1.0$''$ slit width.  We also obtained blue-side spectra with the 600/4000 grism, although those data are not used in this work. The average seeing was  $\sim$0.6$''$  on the first night and $\sim$ 1.0$''$ on the second night.  The targets were divided roughly equally between two masks.  For each mask we obtained 18 exposures of 900s each, giving a total exposure time of 4.5 hours.

The data were reduced using a custom software package developed for faint-object spectroscopy.  Individual exposures were flat-fielded and optimally sky-subtracted in the un-rectified frame following \citet{Kelson2003}.  A combination of alignment star traces and slit edges was used to compute a spatial solution for each exposure. A single rectified two-dimensional spectrum for each object was then extracted simultaneously from all exposures.   Finally, we extracted one-dimensional spectra from the combined 2D spectra using boxcar extraction. The spatial aperture size used for extraction was $2.025 $ arcseconds, which was larger than the typical seeing.  

For some objects we noted small zero-point offsets (up to $\sim 2\times 10^{-20}$ erg cm$^{-2}$ s$^{-1}$ \AA$^{-1}$) in the 1D spectra over extended regions blueward of \lya, which we attribute to small errors in sky subtraction or contamination from foreground continuum objects.  To correct for these effects, we measured the median flux within a one arcsecond window spanning $-2000$~$\mathrm{km~s^{-1}}$ to $-500$~$\mathrm{km~s^{-1}}$ (i.e., blueward) of \lya\ at the systematic redshift of the object, and subtracted it  from the 2D spectrum over all wavelengths before re-extracting the 1D spectrum.   In some cases this procedure also helped remove obvious continuum flux from foreground objects (e.g., ID 13503; Figure~\ref{fig:nondetection2}).

We found that our formal error estimates closely reflected the pixel-to-pixel scatter in the 1D spectrum except over wavelengths affected by strong sky emission lines.  We remedied this by increasing the errors in skyline regions using the following procedure.  We first measured the minimum error within  $\pm 2000$~$\mathrm{km~s^{-1}}$ of \lya\ at the systematic redshift.  The excess error above this minimum was then increased by a factor of two for all pixels.  This procedure effectively increases the errors around skylines by a factor of two while leaving the error at other wavelengths relatively unchanged.

A representative sample of our LRIS spectra is shown in Figure \ref{fig:representativesamples}.  The remainder are included in Appendix \ref{subsec:fullsamplespectra}.  \lya\ line fluxes were measured by integrating the 1D spectra in a window from 0 to $500$~$\mathrm{km~s^{-1}}$ redward of the systematic redshift measured from \Oiii.  The choice of window is motivated by the maximum velocity extent of \lya\ seen in individual objects (e.g., ID 5955 and ID 1233) and the apparent absence of blue peaks. The \lya\ line fluxes are corrected for the underlying continuum using estimates derived from NIRCam {\it JWST} F115W and F200W photometry \citep{Kashino2023}. A power-law continuum of the form $f_\lambda \propto \lambda^{\beta}$ was fit to the photometry, and subtracted from the observed Ly$\alpha$ fluxes. The \lya\ measurements are summarized in Table \ref{tab:summary}. Our median 5$\sigma$ sensitivity is $2.9\times 10^{-17}$ ${\rm erg\, s^{-1}\, cm^{-2}}$ over $5.3 < z <6.3$; however, residuals from strong skylines significantly decrease our sensitivity for some objects, as reflected in the flux uncertainties. We note that Ly$\alpha$ measurements are not corrected for slit losses.

In total, we detect \lya\ emission in 10 (17) galaxies out of 46 with at least 5$\sigma$ (3$\sigma$) confidence.  The spatial and redshift distribution of the detections is illustrated in Figure~\ref{fig:3d-figure}.  As shown below, our overall detection rate is consistent with other recent searches for \lya\ from \Oiii-selected galaxies \citep{Tang2024a}. We note, however, that our detection rate varies significantly between groups.  Some of this is due to observational effects.  For example, our sensitivity to \lya\ is lower for the $z \simeq 5.58$ and $z \simeq 5.99$  groups due to strong skyline residuals. Redshifts with strong skyline residuals (corresponding to wavelengths where the LRIS flux errors are greater than 1.4 times the median) are shown in the lower panel of Figure \ref{fig:3d-figure}. There are three cases, however, for which the fraction of detected \lya\ seems to differ significantly from the mean. The group of five galaxies at $z \simeq 5.94$ show effectively \lya\ emission.  This is a small group for which the lack of \lya\ may be a statistical fluctuation.  The \lya\ fraction also appears to be unusually low, however, in the larger group of 15 galaxies at $z \simeq 6.19$.  Finally, our detection rate is notably {\it higher} than average in the pair of smaller groups at $z \simeq 5.73$ and $5.78$.  Our main focus of this paper is the $z \simeq 6.19$ group, but we return to the objects at $z \simeq 5.73$--5.78 in Section \ref{subsection:line_of_sight}.

\renewcommand{\arraystretch}{1.5} 

\begin{table*}
    \centering
    \begin{adjustbox}{max width=\textwidth}
    \begin{tabular}{|c|c|c|c|c|c|l|c|c|}
        \hline
        ID & RA & Dec & $z_{\text{spec}}$ & $M_{\text{UV}}$ (mag) & $\log_{10}(\text{L}_{\mathrm{[O\,\textsc{iii}]}} / \mathrm{erg\,s}^{-1})$ &  Ly$\alpha$ Flux& $W$ (Ly$\alpha$)&Note \\
                 & deg & deg &  &  (mag) & &$\times 10^{-18}$ erg cm$^{-2}$ s$^{-1}$ & \AA &  \\ \hline \hline

5891 & 15.0554 & 28.0570 & 5.338 & $-20.32 \pm 0.20$ & $42.45 \pm 0.08$ & $7.4 \pm 6.7$ & $25.8_{-23.3}^{+27.3}$&----\\
9327 & 15.0758 & 28.0129 & 5.363 & $-18.27 \pm 0.15$ & $42.10 \pm 0.12$ & $5.6 \pm 3.0$ & $244.5_{-136.7}^{+201.1}$&---- \\
6228 & 15.0550 & 28.0552 & 5.508 & $-19.67 \pm 0.16$ & $41.96 \pm 0.14$ & $1.0 \pm 8.9$ & $10.8_{-95.8}^{+97.4}$&---- \\
8768 & 15.0461 & 28.0555 & 5.570 & $-19.77 \pm 0.07$ & $42.14 \pm 0.07$ & $10.1 \pm 6.9$ & $47.3_{-32.2}^{+33.0}$&---- \\
12083 & 15.0151 & 28.0521 & 5.573 & $-19.88 \pm 0.09$ & $42.47 \pm 0.05$ & $-2.9 \pm 6.4$ & $-15.1_{-33.5}^{+32.9}$&Group@5.58 \\
12410 & 15.0112 & 28.0561 & 5.575 & $-19.04 \pm 0.19$ & $42.46 \pm 0.07$ & $12.3 \pm 6.8$ & $123.2_{-69.5}^{+88.5}$&Group@5.58 \\
12314 & 15.0112 & 28.0574 & 5.575 & $-20.67 \pm 0.08$ & $43.15 \pm 0.02$ & $0.0 \pm 6.5$ & $0.1_{-17.0}^{+16.8}$&Group@5.58 \\
12919 & 15.0091 & 28.0637 & 5.579 & $-18.78 \pm 0.21$ & $42.07 \pm 0.12$ & $3.7 \pm 3.7$ & $61.3_{-61.7}^{+78.3}$ &Group@5.58\\
5955 & 15.0873 & 28.0129 & 5.659 & $-19.84 \pm 0.12$ & $42.65 \pm 0.04$ & $65.2 \pm 2.7$ & $303.2_{-37.8}^{+45.9}$ &----\\ 
1233 & 15.0480 & 28.0911 & 5.718 & $-17.97 \pm 0.38$ & $42.17 \pm 0.11$ & $16.9 \pm 2.0$ & $346.6_{-134.4}^{+314.5}$&Group@5.73\\ 
82 & 15.0572 & 28.0859 & 5.736 & $-19.15 \pm 0.26$ & $42.34 \pm 0.09$ & $15.4 \pm 2.1$ &  $118.8_{-36.5}^{+64.6}$&Group@5.73 \\ 
8005 & 15.0368 & 28.0712 & 5.772 & $-18.63 \pm 0.15$ & $42.13 \pm 0.05$ & $6.7 \pm 2.0$ & $105.2_{-34.7}^{+42.8}$&Group@5.78 \\
982 & 15.0640 & 28.0711 & 5.775 & $-19.65 \pm 0.09$ & $42.42 \pm 0.05$ & $8.1 \pm 2.1$ & $55.3_{-15.0}^{+16.2}$&Group@5.78 \\ 
5052 & 15.0460 & 28.0741 & 5.777 & $-20.23 \pm 0.11$ & $42.42 \pm 0.08$ & $18.2 \pm 2.1$& $72.7_{-12.4}^{+15.1}$&Group@5.78 \\ 
2285 & 15.0571 & 28.0739 & 5.777 & $-21.57 \pm 0.05$ & $43.03 \pm 0.04$ &  $19.4 \pm 2.2$ & $23.1_{-2.8}^{+2.9}$&Group@5.78\\ 
19002 & 15.0438 & 28.0476 & 5.908 & $-18.75 \pm 0.11$ & $41.93 \pm 0.08$ & $9.7 \pm 8.1$ & $174.0_{-144.5}^{+154.3}$&---- \\
1591 & 15.0846 & 28.0402 & 5.925 & $-19.23 \pm 0.17$ & $42.37 \pm 0.09$ & $-7.7 \pm 8.2$ &  $-254.8_{-491.8}^{+268.8}$&Group@5.94 \\
2035 & 15.0882 & 28.0329 & 5.928 & $-19.77 \pm 0.13$ & $42.59 \pm 0.04$ & $-4.2 \pm 8.3$ &$-25.6_{-52.4}^{+50.7}$&Group@5.94 \\
4634 & 15.0702 & 28.0438 & 5.942 & $-19.79 \pm 0.11$ & $42.74 \pm 0.04$ & $5.7 \pm 3.5$ & $43.7_{-26.2}^{+28.6} $&Group@5.94\\ 
8200 & 15.0603 & 28.0387 & 5.943 & $-18.12 \pm 0.14$ & $42.49 \pm 0.03$ & $4.4 \pm 4.2$ & $189.2_{-180.1}^{+209.6}$&Group@5.94 \\
31 & 15.0816 & 28.0535 & 5.944 & $-20.95 \pm 0.07$ & $43.26 \pm 0.06$ & $-4.0 \pm 4.4$ & $-10.5_{-11.7}^{+11.6}$&Group@5.94 \\
4741 & 15.04895 & 28.0729 & 5.987 & $-21.56 \pm 0.07$ & $43.11 \pm 0.02$ & $-10.1 \pm 5.7$ & $-16.4_{-9.4}^{+9.3}$&Group@5.99 \\
6343 & 15.0404 & 28.0743 & 5.987 & $-19.09 \pm 0.15$ & $42.53 \pm 0.08$ & $6.0 \pm 5.7$ & $69.0_{-65.7}^{+71.7}$&Group@5.99 \\
3618 & 15.0432 & 28.0855 & 5.990 & $-20.78 \pm 0.08$ & $42.48 \pm 0.08$ & $16.6 \pm 5.3$ & $42.1_{-13.8}^{+14.6}$&Group@5.99 \\
4396 & 15.0505 & 28.0714 & 5.990 & $-20.13 \pm 0.07$ & $42.51 \pm 0.04$ & $58.5 \pm 5.3$& $236.9_{-25.8}^{+27.7}$&Group@5.99 \\ 
6898 & 15.0649 & 28.0388 & 6.014 & $-20.17 \pm 0.12$ & $42.41 \pm 0.06$ & $7.7 \pm 2.4$& $43.6_{-14.8}^{+19.1}$&---- \\ 
13635 & 15.0600 & 27.9975 & 6.175 & $-18.99 \pm 0.18$ & $42.56 \pm 0.04$ & $2.8 \pm 2.0$ & $40.5_{-28.6}^{+35.0}$&Group@6.19 \\
9209 & 15.0825 & 28.0052 & 6.177 & $-21.68 \pm 0.07$ & $43.42 \pm 0.04$ & $-3.9 \pm 2.0$ & $-5.2_{-2.8}^{+2.7}$&Group@6.19 \\
7038 & 15.0893 & 28.0051 & 6.179 & $-20.02 \pm 0.14$ & $42.35 \pm 0.07$ & $10.2 \pm 2.1$ & $65.1_{-16.7}^{+22.0}$&Group@6.19 \\
10225 & 15.0778 & 28.0072 & 6.179 & $-20.03 \pm 0.12$ & $43.36 \pm 0.06$ & $0.2 \pm 2.1$ & $1.6_{-15.3}^{+15.3}$&Group@6.19 \\
19597 & 15.0671 & 28.0189 & 6.182 & $-18.37 \pm 0.09$ & $42.26 \pm 0.07$ & $3.4 \pm 2.1$ & $78.0_{-48.4}^{+53.2}$&Group@6.19 \\
4304 & 15.0851 & 28.0255 & 6.183 & $-20.49 \pm 0.08$ & $42.90 \pm 0.03$ & $-1.0 \pm 2.2$ &  $-4.6_{-10.0}^{+10.1}$&Group@6.19 \\
10020 & 15.0732 & 28.0139 & 6.184 & $-19.70 \pm 0.11$ & $42.47 \pm 0.04$ & $6.9 \pm 2.1$ & $64.5_{-20.6}^{+24.1}$&Group@6.19 \\
13503 & 15.0766 & 28.0230 & 6.184 & $-20.33 \pm 0.15$ & $42.37 \pm 0.06$ & $2.1 \pm 2.1$ & $8.8_{-9.1}^{+10.9}$&Group@6.19 \\
11496 & 15.0563 & 27.9933 & 6.185 & $-20.01 \pm 0.14$ & $42.83 \pm 0.05$ & $29.4 \pm 2.1$ & $174.0_{-34.3}^{+49.3}$&Group@6.19 \\ 
11989 & 15.0538 & 27.9965 & 6.186 & $-19.12 \pm 0.32$ & $42.54 \pm 0.08$ & $0.7 \pm 2.1$ & $8.2_{-45.7}^{+97.0}$&Group@6.19 \\
4738 & 15.0630 & 28.0526 & 6.187 & $-19.04 \pm 0.11$ & $42.47 \pm 0.04$ & $20.9 \pm 2.2$ & $330.4_{-54.4}^{+67.3}$&Group@6.19 \\ 

        \hline
    \end{tabular}
    \end{adjustbox}
\caption{\centering Information and Ly$\alpha$ properties of  observed galaxies. IDs, spectroscopic redshifts,  $M_{\text{UV}}$, and $\log_{10}( \text{L}_{\mathrm{[O\ III]}}/ \text{erg s}^{-1})$ are from \citet{Kashino2023} and \citet{Matthee2023}. Using
photometry data from \citet{Matthee2023} to estimate the continuum flux density, we
report the median and marginalized 68 percent confidence intervals for the Ly$\alpha$ $W$.}
    \label{tab:summary}

\end{table*}

\begin{table*}
    \centering
    \begin{adjustbox}{max width=\textwidth}
    \begin{tabular}{|c|c|c|c|c|c|l|c|c|}
        \hline
        ID & RA & Dec & $z_{\text{spec}}$ & $M_{\text{UV}}$ (mag) & $\log_{10}(\text{L}_{\mathrm{[O\,\textsc{iii}]}} / \mathrm{erg\,s}^{-1})$

 &  Ly$\alpha$ Flux& $W$ (Ly$\alpha$)&Note \\
                 & deg & deg &  &  (mag) & &$\times 10^{-18}$ erg cm$^{-2}$ s$^{-1}$ &  \AA &\\ \hline \hline
10146 & 15.0683 & 28.0208 & 6.187 & $-20.69 \pm 0.08$ & $42.96 \pm 0.11$ & $7.8 \pm 2.1$ & $26.8_{-7.5}^{+7.9}$&Group@6.19 \\
7979 & 15.0582 & 28.0424 & 6.188 & $-19.53 \pm 0.11$ & $42.39 \pm 0.05$ & $-1.3 \pm 2.1$ & $-10.3_{-18.3}^{+17.7}$&Group@6.19\\
8913 & 15.0786 & 28.0112 & 6.193 & $-19.74 \pm 0.17$ & $42.61 \pm 0.10$ & $6.4 \pm 2.8$ & $49.9_{-23.3}^{+31.0}$&Group@6.19 \\                
16346 & 15.0710 & 27.9953 & 6.194 & $-20.07 \pm 0.15$ & $42.89 \pm 0.04$ & $2.3 \pm 3.3$ & $15.9_{-23.2}^{+26.1}$ &Group@6.19\\
12821 & 15.0424 & 28.0173 & 6.203 & $-19.01 \pm 0.22$ & $42.29 \pm 0.08$ & $19.0 \pm 9.9$ & $141.9_{-78.0}^{+99.3}$&---- \\
16842 & 15.0406 & 28.0415 & 6.206 & $-17.83 \pm 0.28$ & $42.29 \pm 0.04$ & $-6.1 \pm 9.5$ & $-167.8_{-320.3}^{+265.8}$ &----\\
6761 & 15.0712 & 28.0303 & 6.229 & $-18.30 \pm 0.17$ & $42.09 \pm 0.07$ & $-2.3 \pm 4.5$ & $-54.4_{-111.6}^{+104.4}$&---- \\
20317 & 15.0056 & 28.0547 & 6.240 & $-19.66 \pm 0.18$ & $42.19 \pm 0.11$ & $25.4 \pm 2.1$ & $171.7_{-43.0}^{+69.6}$&---- \\ 
11016 & 15.0347 & 28.0190 & 6.247 & $-17.68 \pm 0.36$ & $41.97 \pm 0.08$ & $13.5 \pm 2.3$& $472.3_{-182.3}^{+421.1}$&---- \\ 

        \hline
    \end{tabular}
    \end{adjustbox}
    \contcaption{  }

\end{table*}

\renewcommand{\arraystretch}{1}

\begin{figure*}
    \centering
    \includegraphics[width=0.331\textwidth]{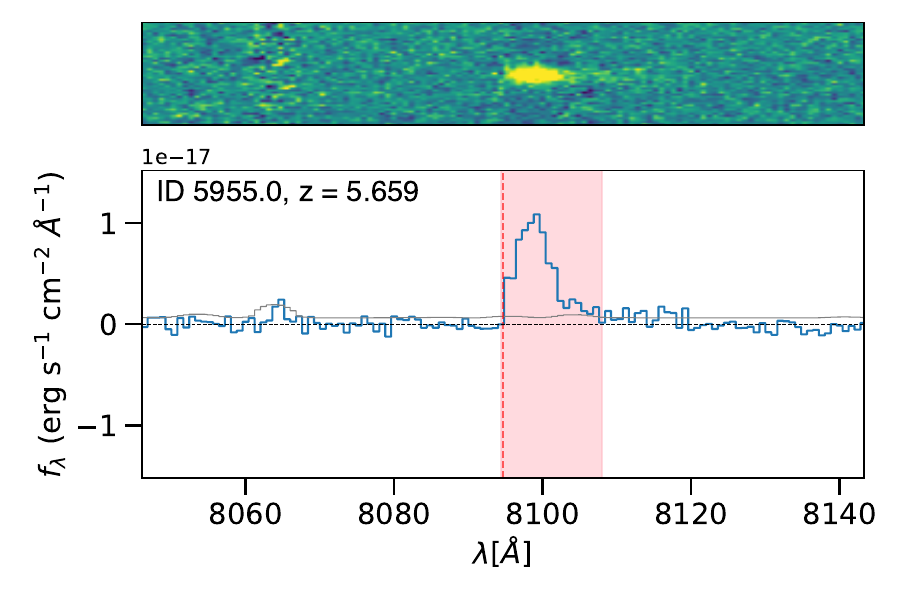}
    \hfill
    \includegraphics[width=0.331\textwidth]{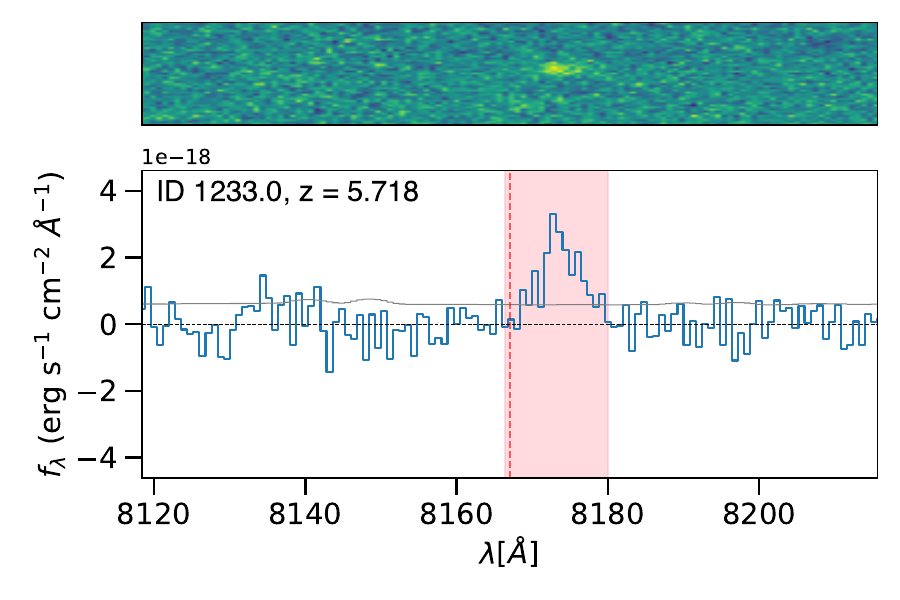}
    \hfill
    \includegraphics[width=0.331\textwidth]{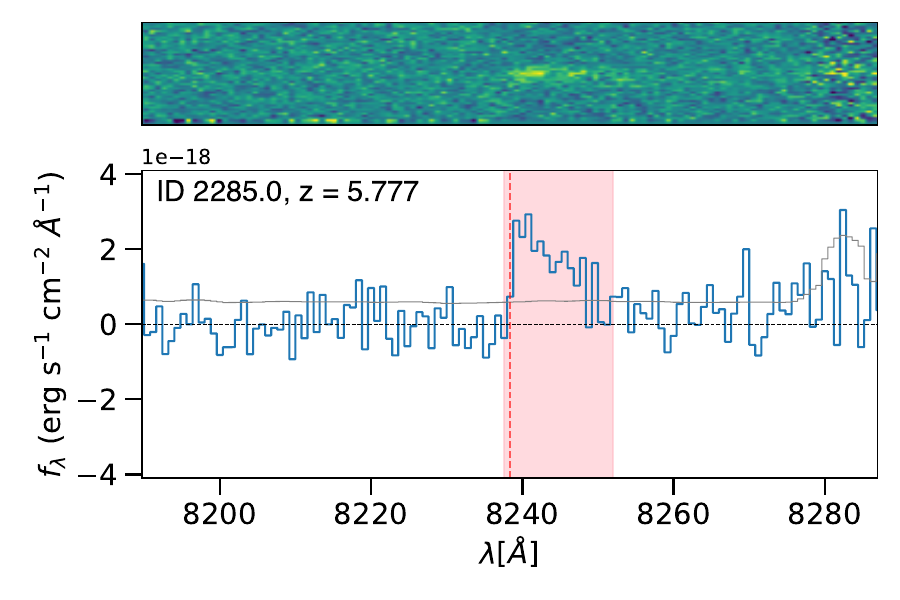}

    \vspace{0.3cm}

    \includegraphics[width=0.331\textwidth]{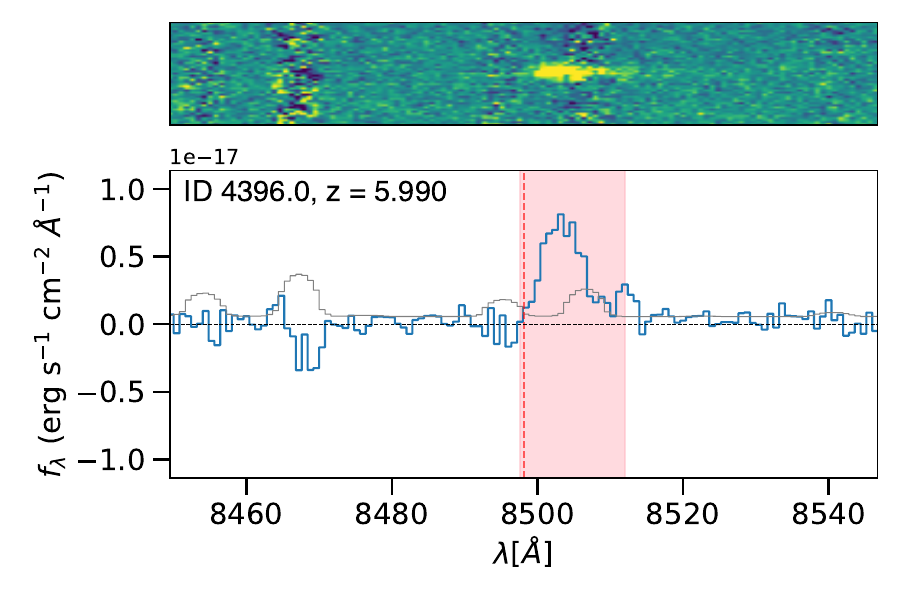}
    \hfill
    \includegraphics[width=0.331\textwidth]{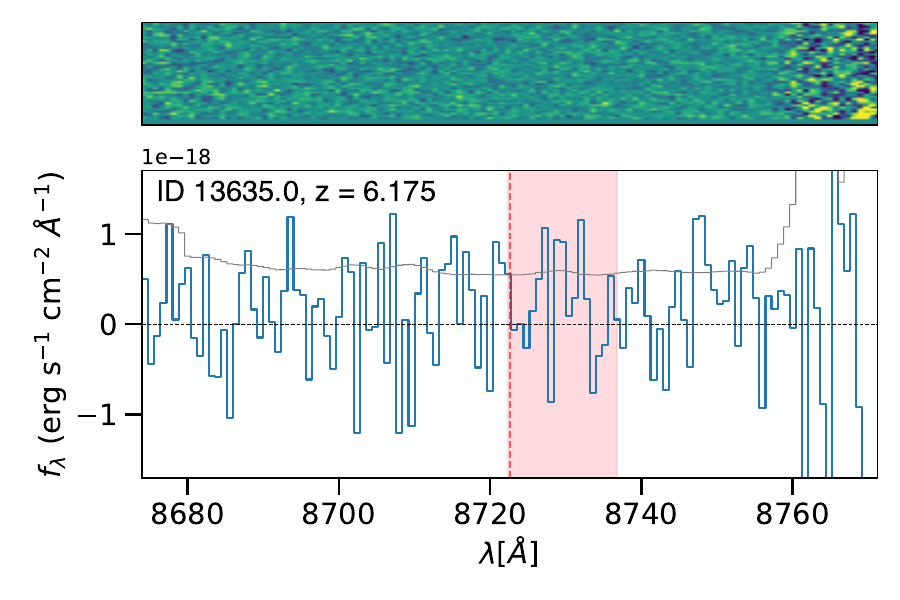}
    \hfill
    \includegraphics[width=0.331\textwidth]{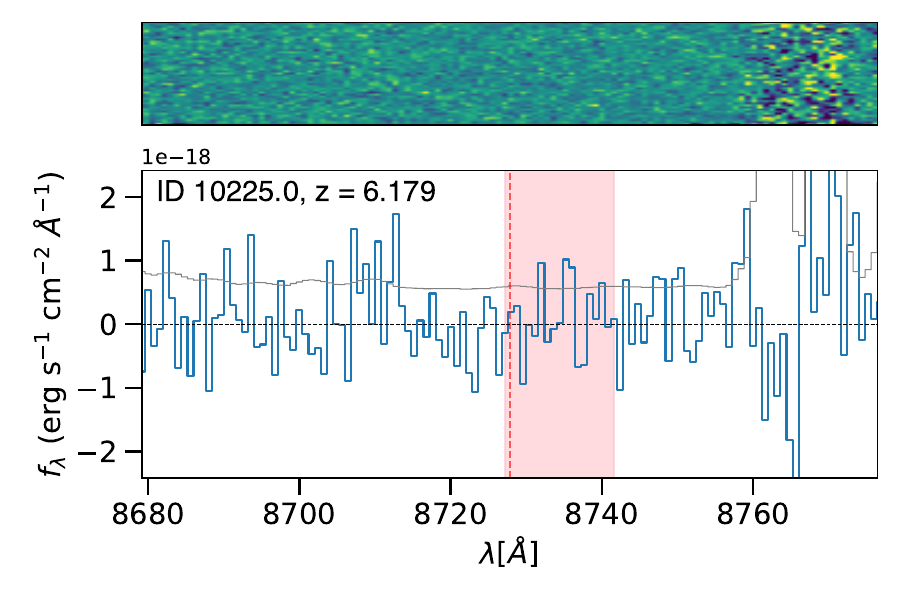}

    \vspace{0.3cm}

    \includegraphics[width=0.331\textwidth]{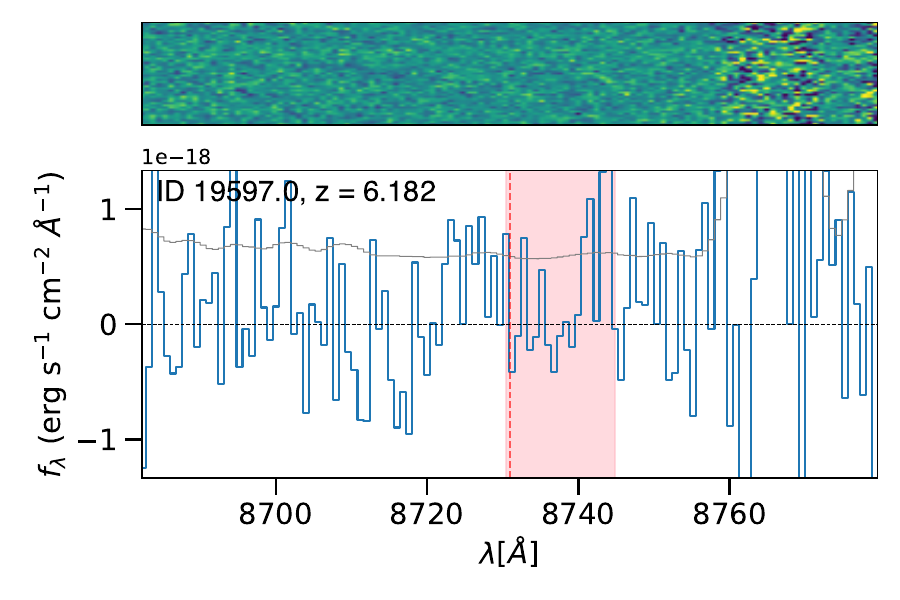}
    \hfill
    \includegraphics[width=0.331\textwidth]{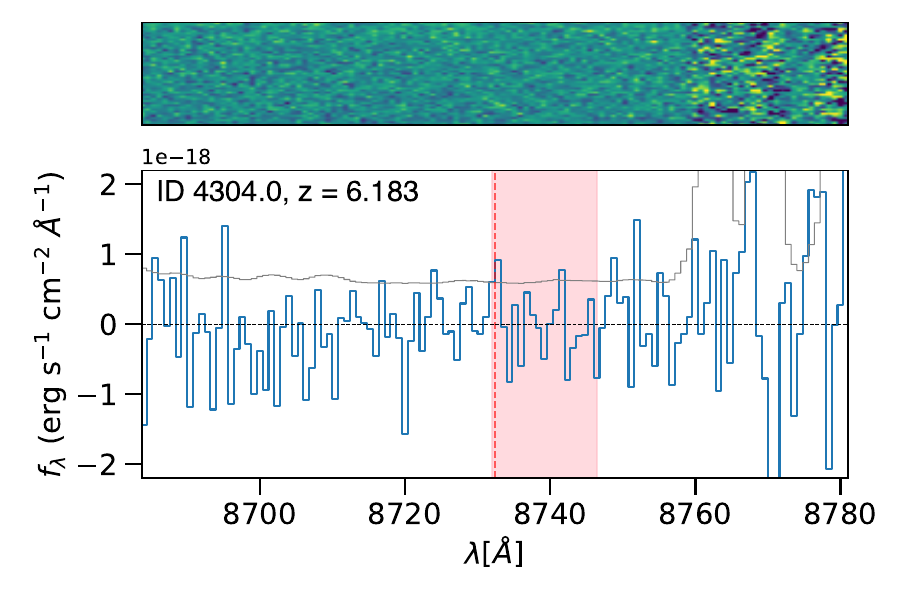}
    \hfill
    \includegraphics[width=0.331\textwidth]{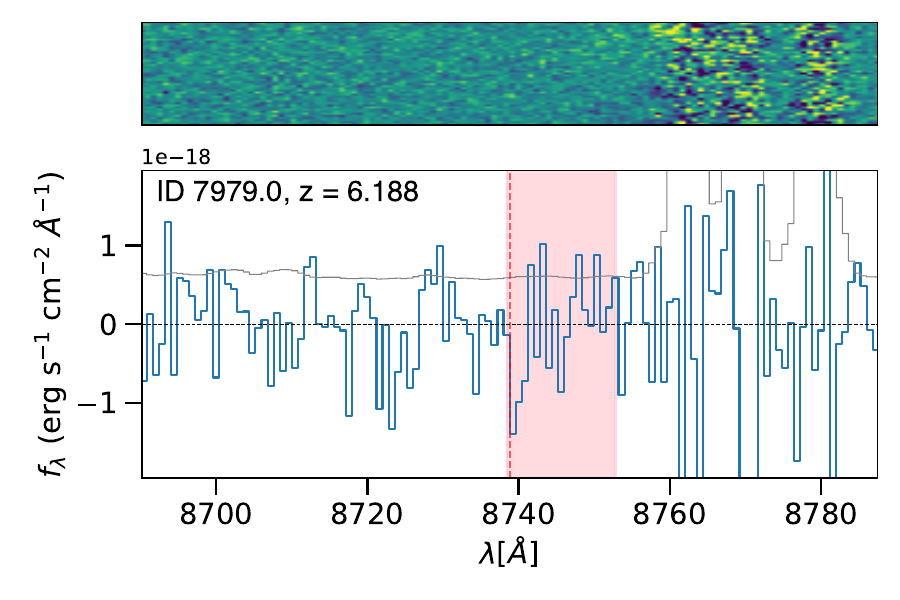}
    
    \hfill
    \hspace{0.331\textwidth} 
    \hfill
    \hspace{0.331\textwidth} 

    \caption{ 1D and 2D emission-line spectra of a representative sample of observed galaxies.  Each panel is centered on the systematic redshift of the galaxy measured from \Oiii\ (red dashed vertical line). We show the 2D  spectrum for each galaxy at the top of each panel. The 1D spectrum (blue histogram) of each emitter in the observed frame is also shown. The uncertainty of  the 1D spectrum is plotted in each grey. Shaded regions indicate the 500~$\mathrm{km~s^{-1}}$ windows used to measure \lya\ fluxes. Spectra for the reminder of our objects are shown in figure \ref{fig:nondetection1}.  }
    \label{fig:representativesamples}
\end{figure*}

\begin{figure*}
    \centering
\includegraphics[width=\textwidth, trim=0 1cm 0 1cm, clip]{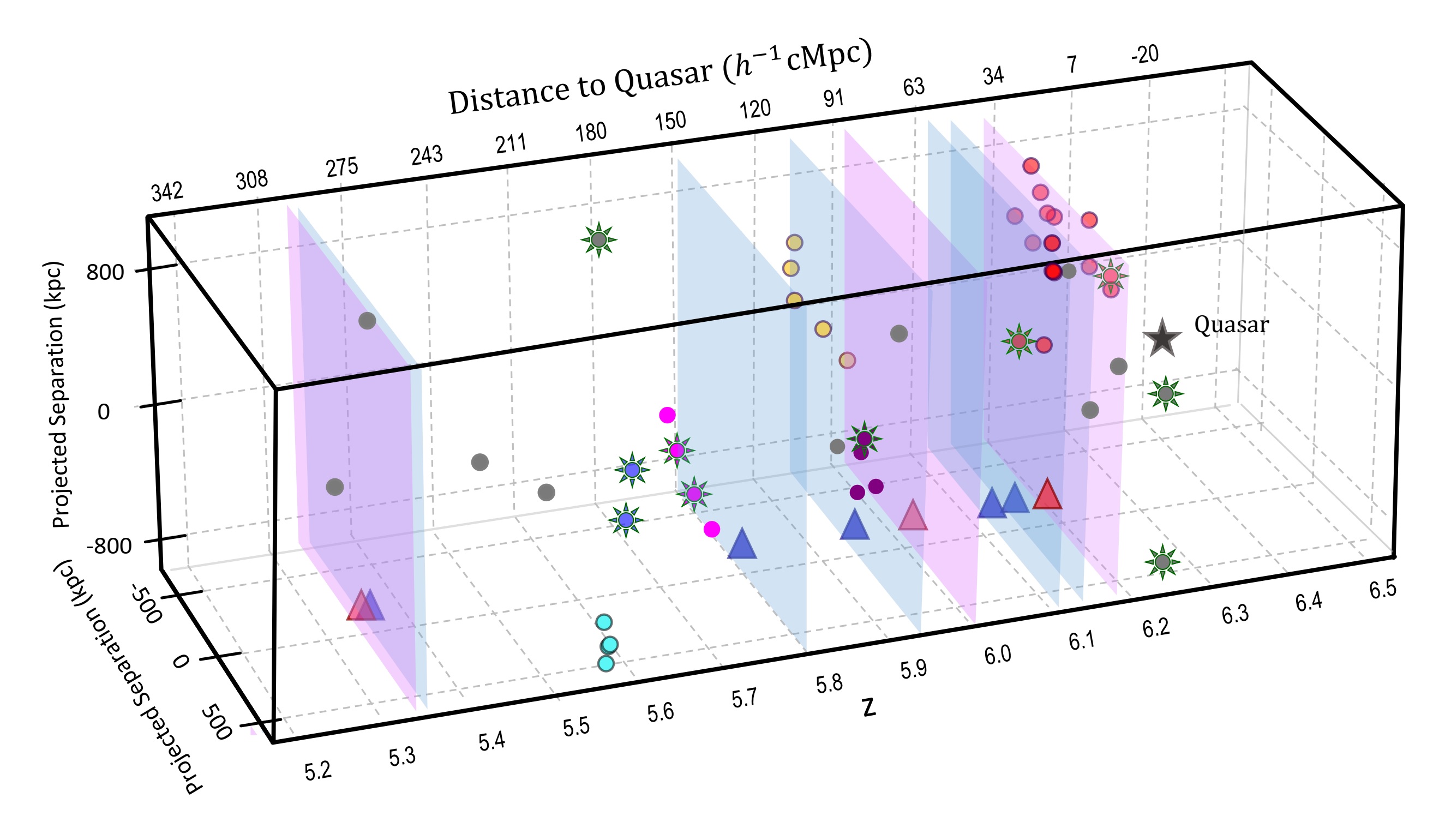}

{\includegraphics[width=\textwidth, trim=0 0cm 0 0cm, clip]{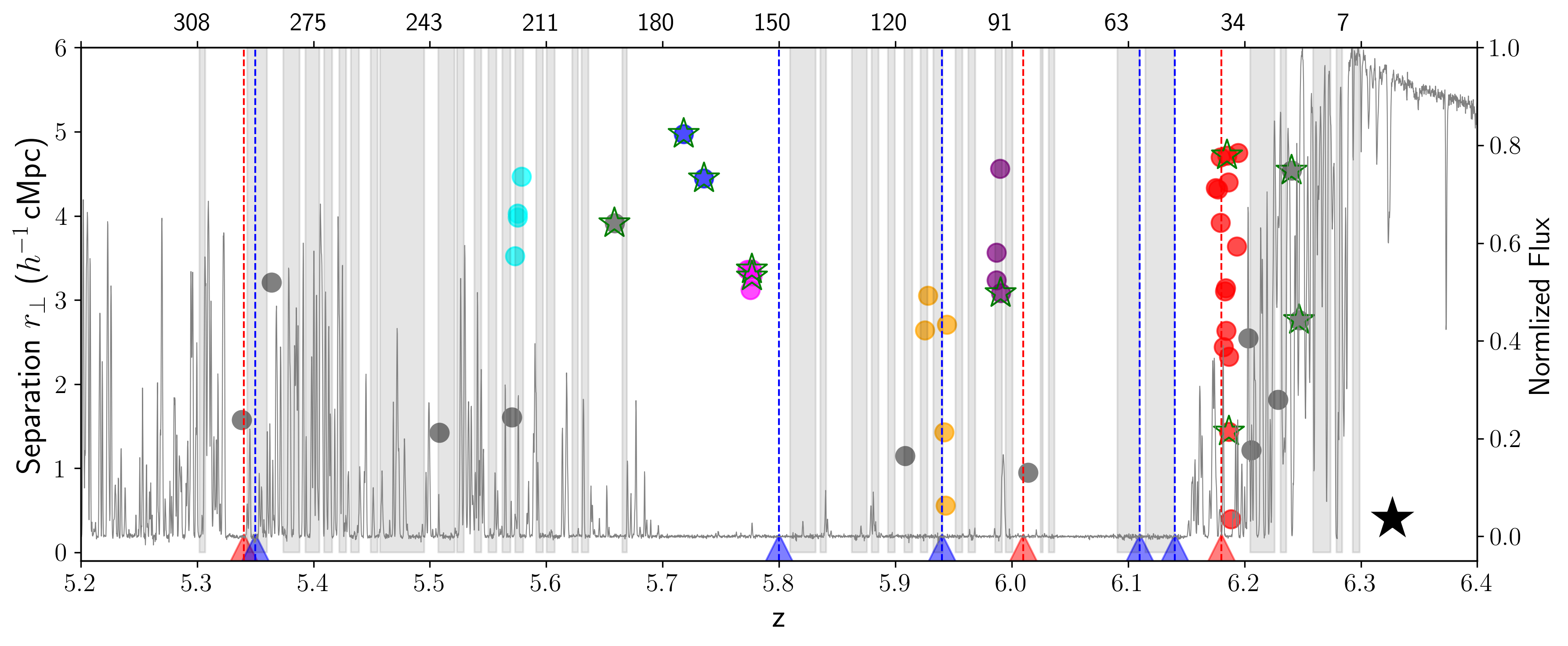}}
    \caption{Illustrations of our sample in two viewing angles. Circles show our targeted galaxies, with colors correposnding to Figure~\ref{fig:sample_spatial_distribution}.  A green star means  that the galaxy is detected in \lya\ at $>$5$\sigma$ confidence. The black star at the end of the box shows the quasar redshift. Upper axes in both figures show the co-moving distance to the quasar. A blue (red) triangle, extended by a light blue (pink) plane, indicates the redshift of a low-ionization (high-ionization) metal line absorber identified in the spectra of the quasar.  Shaded parts in the lower panel show regions affected by sky lines.  }
    \label{fig:3d-figure}
\end{figure*}

\section{Results }
\label{section:results}

\subsection{Global \lya\ equivalent width distribution}
\label{subsec:lyaequivalentwidths}

We first examine the global properties of \lya\ emission in our sample, then turn to variations between groups below. We convert our \lya\ flux measurements to equivalent widths using continuum estimates derived from photometry (see Section~ \ref{subsec:kecklrisdata} for more detail).   The $W$ values are computed by dividing the measured Ly$\alpha$ line fluxes by the continuum flux density determined from the power law at 1216~\AA.  Errors in the equivalent widths are computed using a Monte Carlo approach that includes both uncertainties in the \lya\ fluxes and uncertainties in the {\it JWST}
fluxes.  The results are summarized in Table \ref{tab:summary}.


For the $W$ distribution, we apply a Bayesian framework and assume a log-normal distribution to construct the Ly$\alpha$ $W$ distributions. This distribution is characterized by a set of parameters $\theta = [\mu, \sigma]$, where $\mu$ and $\sigma$ are the mean and standard deviation, respectively, of $\ln{W}$. We use uniform priors for the model parameters, with $\mu$ ranging from 0 to 6 and $\sigma$ ranging from 0.01 to 4 \citep[see also][]{Schenker2014, Endsley2021, Tang2024a}.  

For each set of model parameters $\theta$, the log-normal distribution is
\begin{equation}
\label{equation:lognormal}
p(W|\theta) = \frac{1}{\sqrt{2\pi} \, \sigma  W}  \exp{\left[-\frac{(\ln{W} - \mu)^2}{2\sigma^2}\right]} \, .
\end{equation}
We compute a likelihood for each galaxy by convolving the model distribution with $p_{{\rm obs}, i}(W)$, the full empirical (and typically non-Gaussian)  probability distribution of $W$ for galaxy $i$,
\begin{equation}
p({\rm obs},i |\theta) = \int^{\infty}_{-\infty}  p_{{\rm obs}, i}(W)  \, p(W|\theta)\, dW \, .
\end{equation}
The likelihood for the entire sample is then determined by multiplying the individual likelihoods for each galaxy, 
\begin{equation}
p({\rm obs}|\theta) \propto \prod_{i} p({\rm obs},i|\theta).
\end{equation}
Using Bayes' theorem, we express the posterior probability distribution for the model parameters as
$p(\theta|{\rm obs}) \propto p(\theta)  p({\rm obs}|\theta)$,
where \( p(\theta) \) represents the prior of the model parameters. We first apply the model to the full sample. The result shown on the left side of Figure \ref{fig:contour} is consistent with a recent study by \cite{Tang2024a}.  As discussed below, however, the \lya\ properties of individual groups may differ significantly from one another.

\subsection{A large group with a significant \lya\ deficit?}
\label{subsection:lya_deficit} 

The group of 15 galaxies at $z\simeq 6.19$ is the largest group in our sample. This structure is located $\sim$38 $h^{-1}\,$cMpc in the foreground of the J0100, corresponding in redshift to the end of the quasar's proximity zone (\citealt{Kashino2023}).  Given that the IGM within the proximity zone must be highly ionized, we might expect that the observed \lya\ fraction for this group would be equal to or greater than the global average.  Notably, however,  we have detected \lya\ emission from only two (five) sources at 5$\sigma$ (3$\sigma$) confidence.  We also note that the two detected LAEs are also located at the edges of the group, with one very close to the quasar line of sight.  We return to this point below.

\begin{figure}
    \centering
    \includegraphics[width=.5\textwidth]{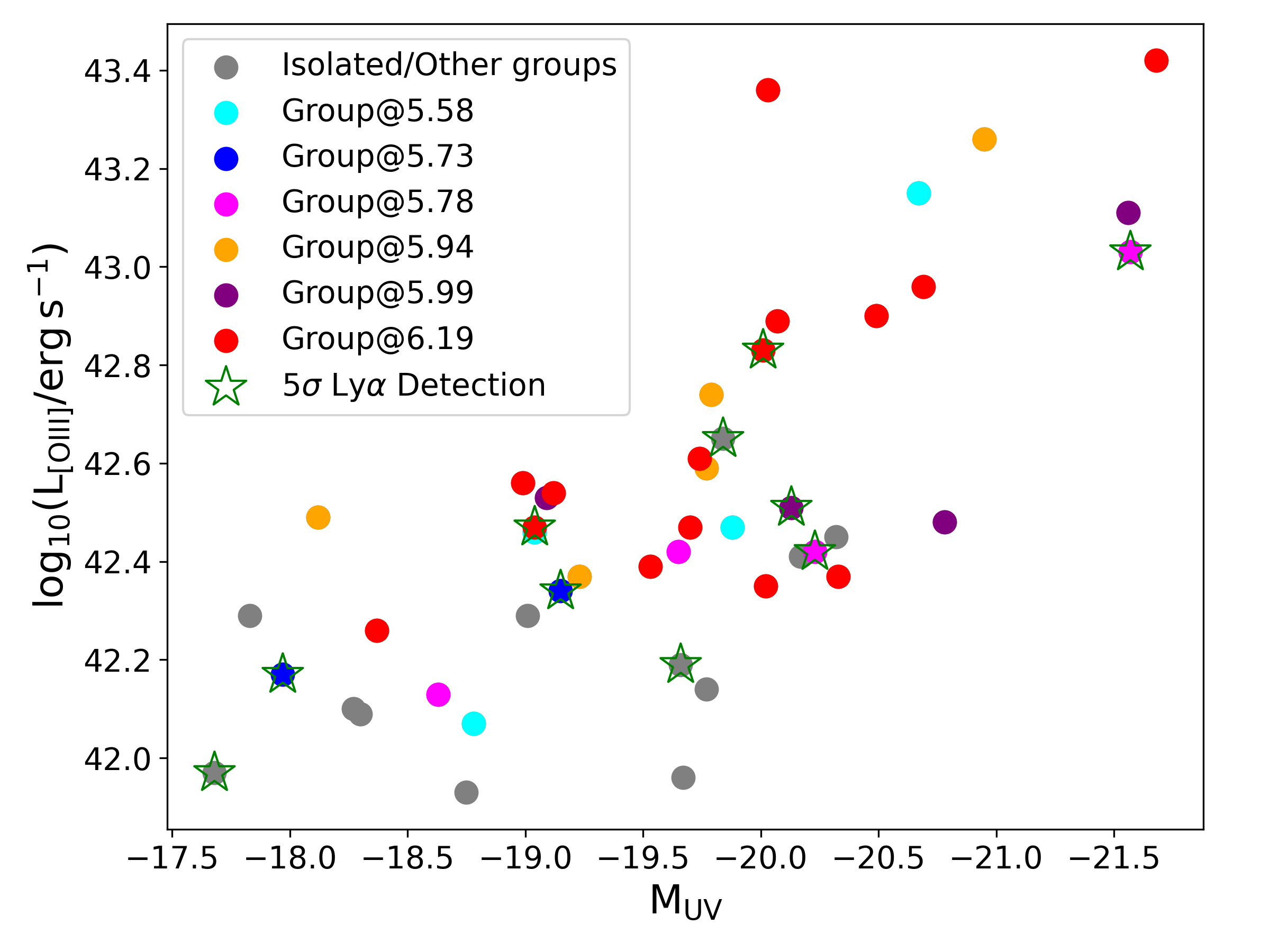}
    \caption{\Oiii\ luminosity vs. UV magnitude for the sources in this study.  Groups are color-coded as in Figure~\ref{fig:sample_spatial_distribution}.  Stars indicate sources with 5$\sigma$ \lya\ detections.}
    \label{fig:uv-o3}
\end{figure}

We note that our \lya\ measurements for the $z \simeq 6.19$ group are not impacted by strong skylines.  Furthermore, the distribution of UV magnitudes and \Oiii\ luminosity of galaxies in this group are broadly consistent with the rest of the sample (see Figure \ref{fig:uv-o3}). We also calculated the mean UV slope for this group ($-1.69 \pm 0.11$) and the rest of the sample ($-1.79 \pm 0.17$) using two filters (F115W and F200W) and found that they are consistent within the 1$\sigma$ errors.  This suggests that the  dust content of the $z \simeq 6.19$ galaxies is not obviously different from the rest of the sample. If the observed \lya\ emission from this group is significantly lower than the remainder of the sample, therefore, it is possible that its \lya\ emission is attenuated by intervening gas.  

\begin{figure*}
    \centering
        \includegraphics[width=0.49\textwidth]{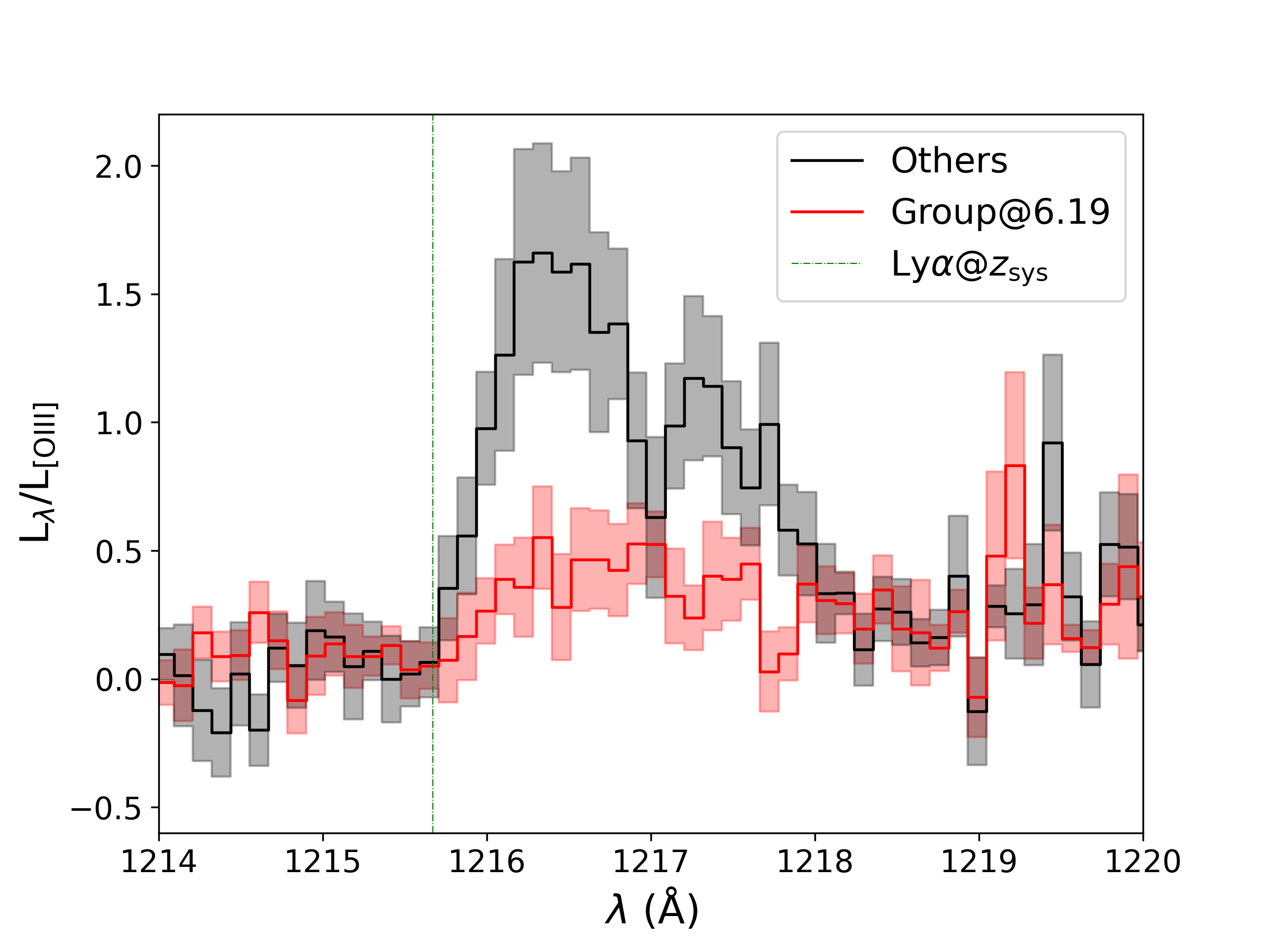}
    \includegraphics[width=0.49\textwidth]{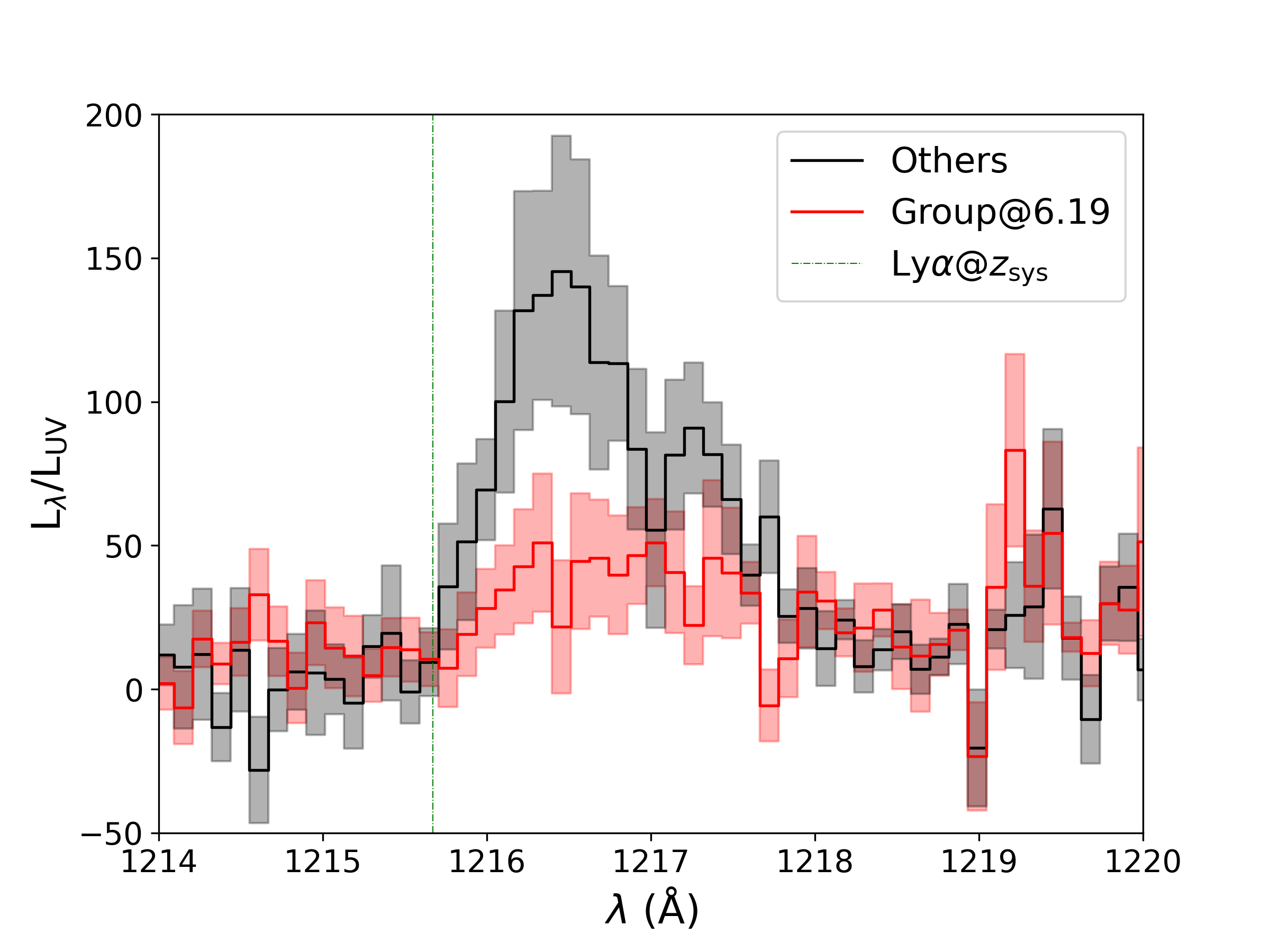}  
    \caption{Mean stacked Keck LRIS spectra near \lya.  Red histograms are for the group at $z \simeq 6.19$, while the black histograms are for the remainder of the sample.  In the left-hand panel the individual spectra are normalized by their \Oiii\ luminosity prior to stacking, while in the right-hand panel they are normalized by their UV luminosity.  Shaded regions show the 1$\sigma$ variation from bootstrap resampling.  The vertical dashed lines correspond to \lya\ at the systemic redshift measured from \Oiii. }
    \label{fig:lyastacked}
\end{figure*}

In order to determine whether the apparent deficit in the $z \simeq 6.19$ group is significant, we compare this group to the rest of the sample using two quantities, the rest-frame \lya\ equivalent width ($W$) distribution, and the ratio of \lya\ and \Oiii\ Luminosities.  As a visual demonstration, we compare the stacked spectra of the $z \simeq 6.19$ group to that of the rest of the sample in Figure~\ref{fig:lyastacked}. To exclude pixels affected  by strong sky lines, we exclude regions from individual spectra where the corresponding flux uncertainty exceeds 1.4 times the median of the full error array for that spectrum. In the left-hand plot, we normalize each individual spectrum by its \Oiii\ luminosity, while in the right-hand plot, they are normalized by their UV luminosities. In both cases the mean \lya\ emission in the $z \simeq 6.19$ group is roughly a factor of three lower than the mean of the other sources. 

\begin{figure*}
    \centering
    \includegraphics[width=0.49\textwidth]{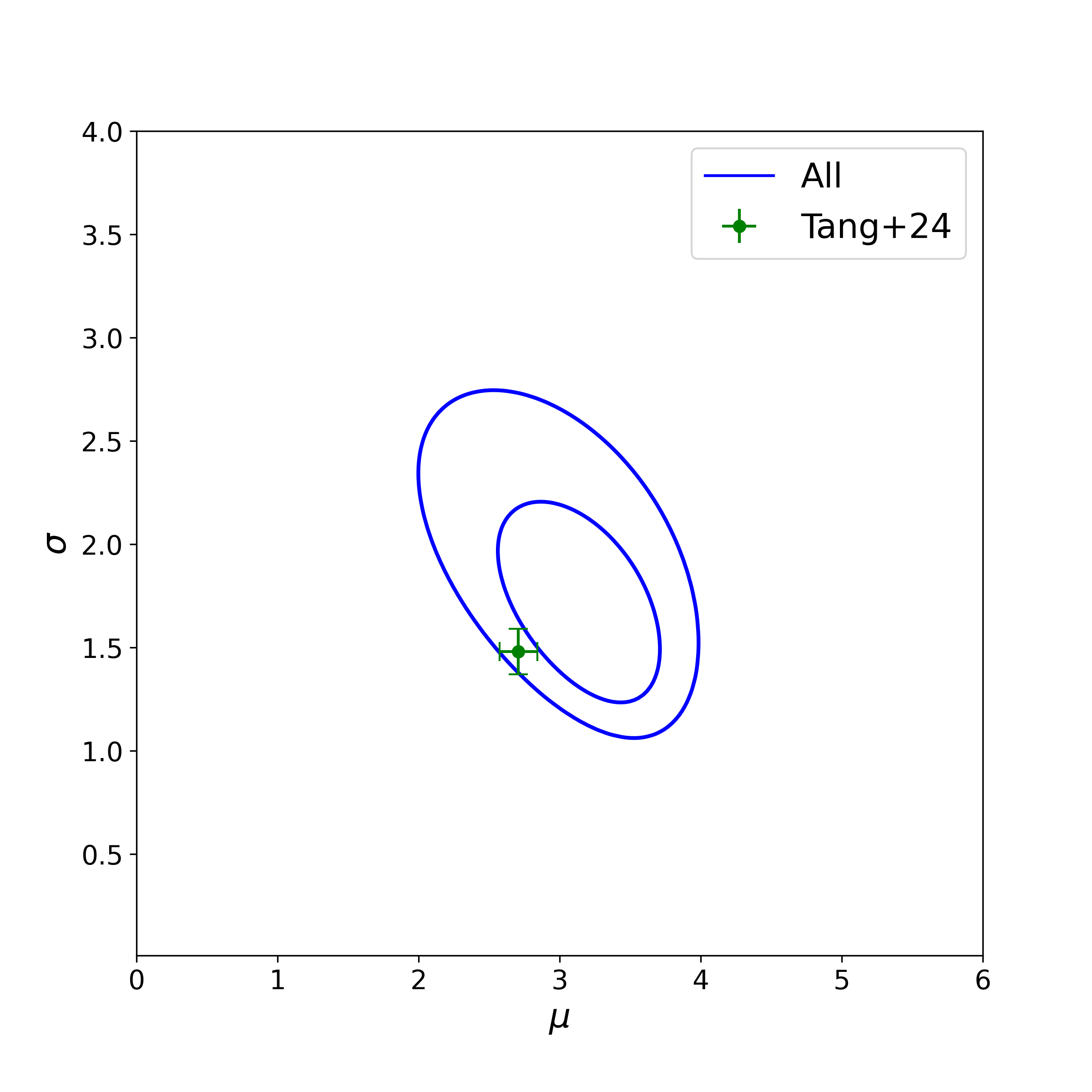}
    \includegraphics[width=0.49\textwidth]{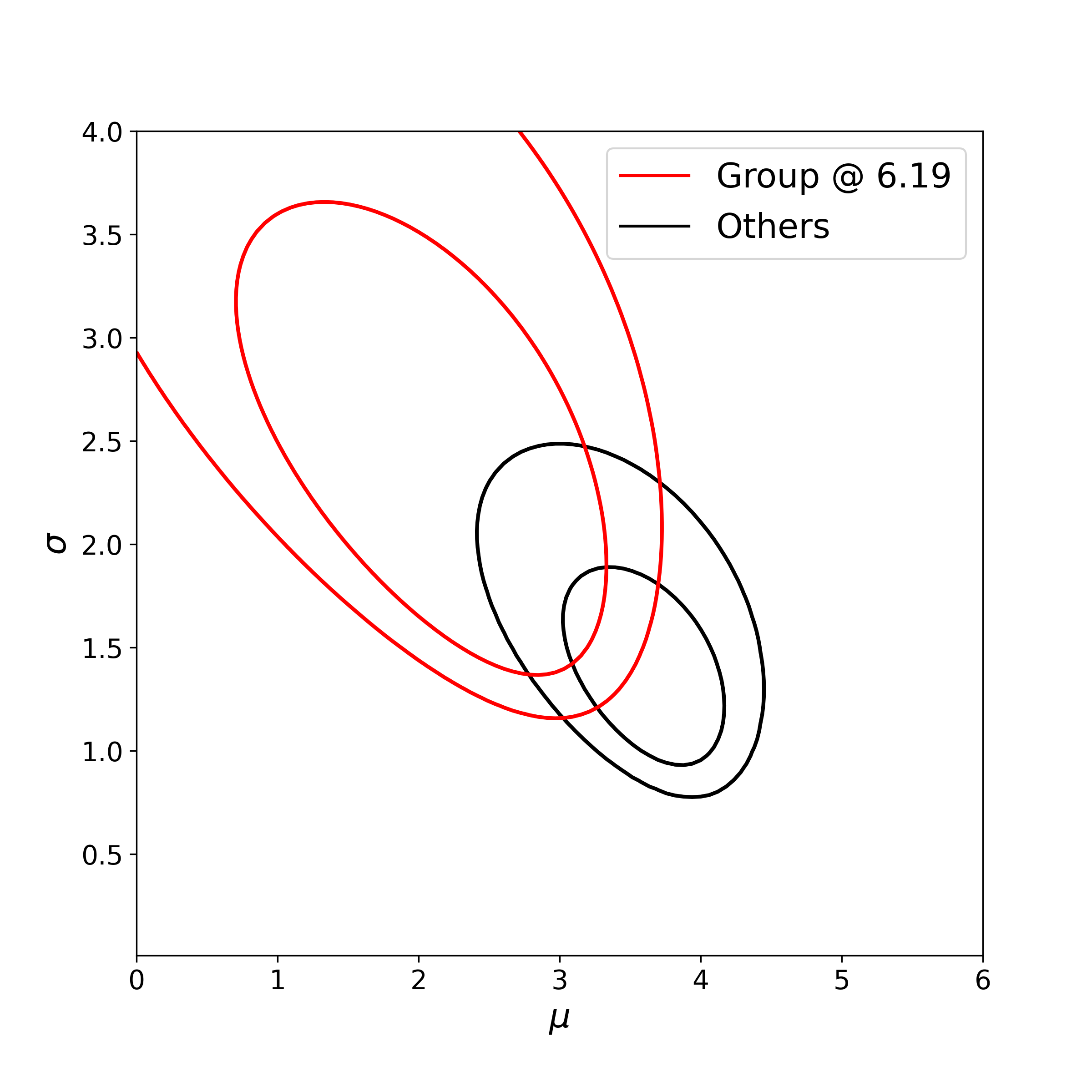}   
    \caption{Constraints on a log-normal distribution of Ly$\alpha$ equivalent widths. The contours show the 68\% and 95\% likelihood bounds for the parameters in Equation~\ref{equation:lognormal}. The left-hand panel shows the result for the full sample, overplotted with results from \citep{Tang2024a}. The right-hand panel shows results for the group at $z \simeq 6.19$ (red) and the remainder of the sample (black).}
    \label{fig:contour}
\end{figure*}

In order to better quantify this apparent difference, we first compare the \lya\ equivalent width distribution of the $z \simeq 6.19$ group to that of the rest of the sample.  We apply the fitting formalism described above to these subsamples separately, with results presented on the right-hand side of Figure \ref{fig:contour}.  There is some evidence that the best-fitting lognormal parameters differ; however, the parameters for the $z \simeq 6.19$ group are poorly constrained due to the small sample size and overlap significantly with those for the rest of the sample. We note that the result does not strongly depend  on the choice of the underlying W distribution model; adopting an exponential functional $p(W) \propto \exp(-W/W_0)$, yields similar conclusions.  As a related test, we adopt the lognormal parameters for the rest of the sample as the parent distribution and calculate the probability that a group of 15 galaxies would randomly have a mean equivalent width as low as the one observed for the $z \simeq 6.19$ group.  For this test, we randomly draw parameters from the distribution shown in Figure~\ref{fig:contour} in a probability-weighted way, and then generate a random sample of 15 $W$ values for each set of parameters.  The randomly generated groups have a mean $W$ that exceeds the $z \simeq 6.19$ value in 70\% of the trials.  This suggests that the $z \simeq 6.19$ group may have a significantly lower mean $W$ than the remainder of the sample, through the evidence from this test is weak.

\begin{figure}
    \centering
    \includegraphics[width=.5\textwidth]{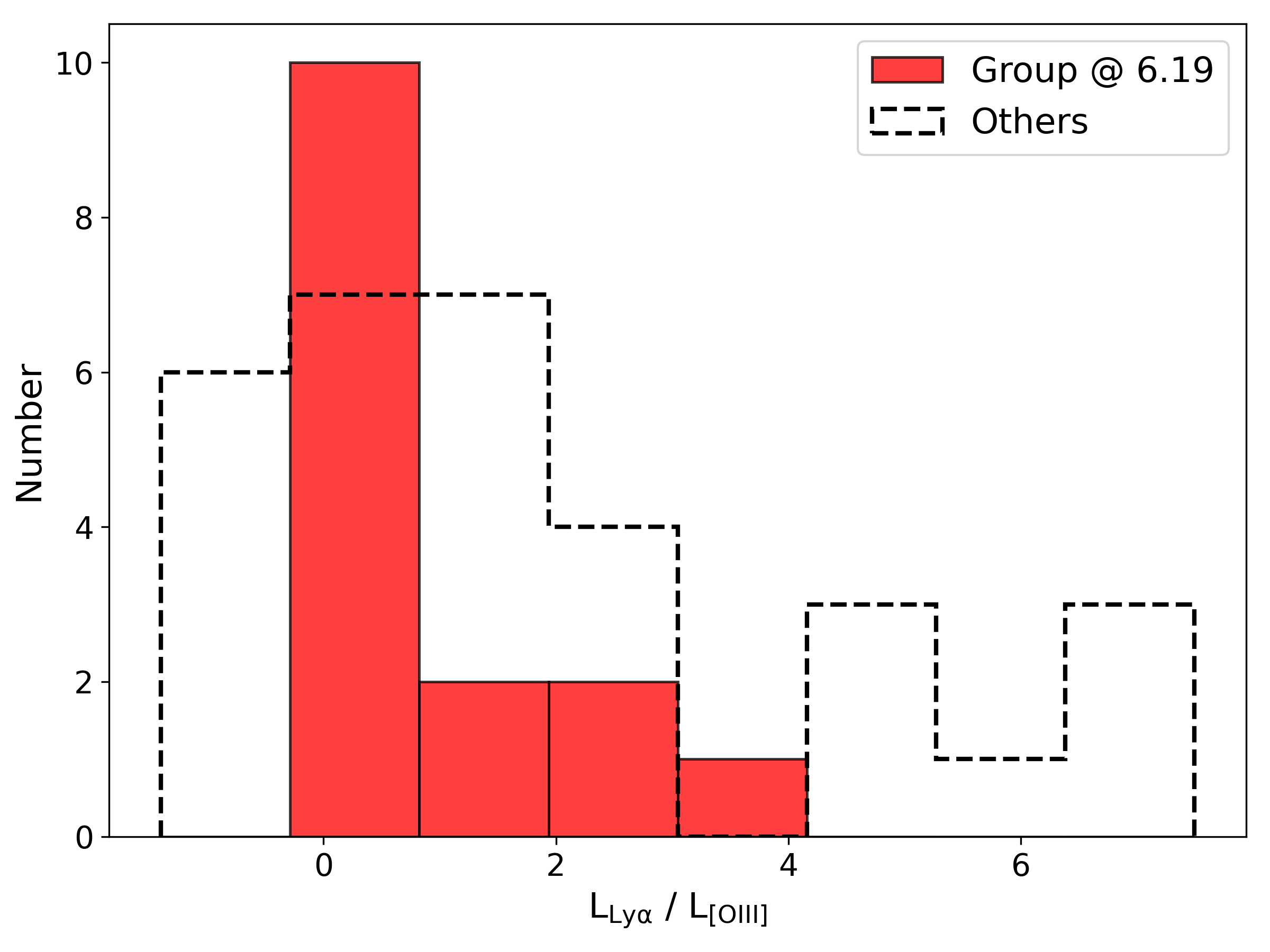}
    \caption{ Distribution of the ratio of Ly$\alpha$ and \Oiii\ luminosities.  Values for the group at 6.19 are shown in red, while the rest of the sample is shown as a dashed line.}
    \label{fig:lya-o3}
\end{figure}

Stronger evidence comes from examining the ratio of \lya\ and \Oiii\ luminosities. In principle, intrinsic \lya\ emission should be strongly correlated with \Oiii\ as both are tracers of star formation.  Figure \ref{fig:lya-o3} presents the distribution of $L_{\rm Ly\alpha} / L_{\rm [O\,III]}$ for both the 15 galaxies at $z \simeq 6.19$ and the remaining 31 galaxies at other redshifts. The other objects include a tail at $L_{\rm Ly\alpha} / L_{\rm [O\,III]} > 4$ that is not present at $z \simeq 6.19$. The mean values are $0.71 _{-0.25} ^{+0.28}$ at $z \simeq 6.19$ and $1.85 _{-0.69} ^{+0.76}$ for the remainder. Here, the 68\% confidence intervals were computed via bootstrap resampling, with the bootstrap draws taken from the full error distribution for each input \lya\ and \Oiii\ measurement.  It is not clear from Figure~\ref{fig:lya-o3} what the underlying functional form of the $L_{\rm Ly\alpha} / L_{\rm [O\,III]}$ distribution should be.  Nevertheless, we can perform an empirical test of whether the two distributions are consistent with one another.  For this we adopt our 31 measurements of $L_{\rm Ly\alpha} / L_{[{\rm O\,III}]}$ outside of the $z \simeq 6.19$ group as our parent distribution, and ask what the probability is of drawing a random set of 15 values from this distribution with a mean value as low as observed at $z \simeq 6.19$.  Here again, individual values are drawn from the full error distribution for each measurement. We find that a random sample has a lower mean in only 6\% of the trials. While the evidence is still modest, therefore, the ratios of \lya\ and \Oiii\ provide some evidence that the mean \lya\ emission measured in the $z \simeq 6.19$ group is significantly lower than that of the rest of the sample.

\subsubsection{Possible causes of the \lya\ deficit}

The comparisons above provide some indication that \lya\ emission from the large group of \Oiii\ emitters at $z \simeq 6.19$ may be attenuated compared to the rest of the sample.  While these test provide only marginal evidence based on the ensemble \lya\ properties, the spatial distirbution of \lya\ in this group is also notable.  Specifically, the two galaxies with \lya\ detected at $>$5$\sigma$ confidence lie at the edges of the structure (Figure~\ref{fig:sample_spatial_distribution}).  One of these, moreover, lies within 50 arcseconds (1.4 $h^{-1}\,$cMpc) of the quasar line sight, and hence, given its redshift, is very likely to lie within the quasar proximity zone even if the quasar ionizing emission is beamed within a cone of $\lesssim$2 degrees.  The fact that there is very little detectable \lya\ from within the interior of this group (galaxies 7038, 10020, and 10146 are detected at only 3$\sigma$ significance) is further reason to suspect that there may be \lya\ attenuation over at least part of the structure.  

We emphasize that the galaxies in the $z \simeq 6.19$ group have UV magnitudes, \Oiii\ luminosities, and UV slopes that are broadly consistent with the full sample (see Figure~\ref{fig:uv-o3} and Section~\ref{subsection:lya_deficit}). Still, the galaxies in this group are on average slightly brighter (Figure~\ref{fig:uv-o3}). We therefore conducted an additional test by restricting the full sample to galaxies with $M_{\mathrm{UV}} < -19$ and recreating the comparisons shown in Figure~\ref{fig:lyastacked}. This analysis produced no significant change in the results, suggesting that even when compared to similarly bright galaxies, those in the group show more attenuated Ly$\alpha$ emission. We also performed a Mann–Whitney U test comparing the \Oiii\ equivalent widths and UV magnitudes between the $z \simeq 6.19$ group and the rest of the sample. The resulting p-values are 0.54 for \Oiii\ equivalent widths and 0.24 for UV magnitude, confirming that there are no statistically significant differences in these properties.

Another useful quantity for comparing the galaxies in these two samples is the \lya\ velocity offset, which can be sensitive to the galaxies' neutral gas content and/or kinematics. Among our 5$\sigma$ detections, the average \lya\ peak velocity offset (measured from the brightest pixel) for the $z \simeq 6.19$ group is  $\sim139$ ~$\mathrm{km~s^{-1}}$, compared to $\sim180$ ~$\mathrm{km~s^{-1}}$ for the rest of the sample. A more robust constraint comes from the velocity offset measured in the stacked spectra, which shows no significant difference in the \lya\ shift relative to the systemic redshift between the two samples.


If \lya\ from this group is unusually weak, it therefore suggests that there may be scattering due to damping wing absorption from intervening neutral hydrogen.  \citet{Heintz2024} presented evidence that a proto-cluster at a redshift of $z = 5.4$ may contain a significant amount of neutral gas based on damping absorption measured in background galaxies.  While the $z \simeq 6.19$ could potentially also host neutral gas, we note that in models where star-forming galaxies dominate the ionizing UV background, galaxy overdensities should tend to be highly ionized.  We therefore consider a scenario in which the damping absorption arises from neutral regions in the foreground of the $z \simeq 6.19$ group.

\begin{figure*}
    \centering
    \includegraphics[width=\textwidth]{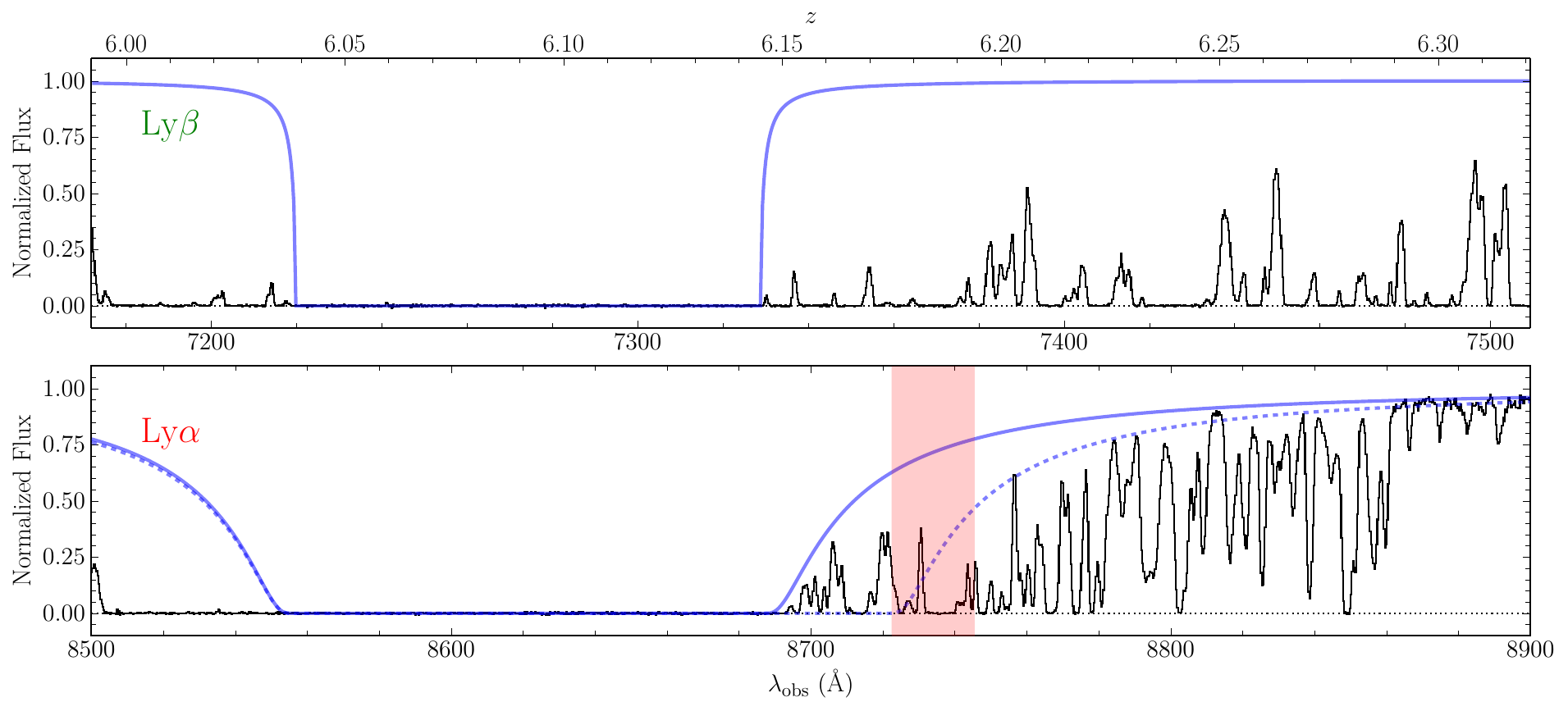}
    \caption{VLT X-Shooter spectrum of J0100+2802 showing transmission in the proximity zone and adjacent absorption trough in Ly$\alpha$ (bottom panel) and Ly$\beta$ (top panel).   Redshifts corresponding to the observed wavelengths are shown along the top axis.  The red shaded region corresponds to the redshift range spanned by the $z\simeq 6.19$ group.   Transmission by a plausible neutral island along the quasar line of sight is shown as a blue continuous line.  Along the line of sight, the red edge of this island would be $\sim$13 $h^{-1}\,$cMpc from the group at $z \simeq 6.19$.  At that separation the island would only attenuate \lya\ transmission at $z=6.19$ by $\sim$25\%.  In contrast, if the red edge of the island is only $\sim$5 $h^{-1}\,$cMpc from the group (which itself is transversely offset from the quasar line of sight by $\sim$3 $h^{-1}\,$cMpc), then the attenuation would increase to $\sim$60\% (dashed line).}
    \label{fig:dampingwing}
\end{figure*}

A neutral island along the line of sight could attenuate \lya\ from the $z \simeq 6.19$ group, and would also be expected to produce strong absorption in the spectrum of the quasar.    Indeed, the X-Shooter spectrum of Figure~\ref{fig:dampingwing} shows complete absorption in both \lya\ and \lyb, as well as all higher-order Lyman lines, from the edge of the proximity zone at $z \simeq 6.14$ down to $z \simeq 6.04$.  Following \citet{Becker2024}, we use the \lya\ and \lyb\ transmission peaks to calculate the largest mean-density neutral island that could exist within this dark gap.  The solid blue line in Figure~\ref{fig:dampingwing} shows the damping wing profile from a 30.0 $h^{-1}\,$cMpc island with its edge located 50.5 $h^{-1}\,$cMpc from the quasar, and so $\sim$13 $h^{-1}\,$cMpc along the line of sigh from the group at $z \simeq 6.19$.  The \lya\ attenuation from such an island would only be $\sim$25\% at the group redshift, smaller than what we observe.  In order to produce a factor of three attenuation, the edge of the island would have to lie $\sim$5 $h^{-1}\,$cMpc from the group.  Increasing the island length by this amount along the quasar line of sight would violate the observed transmission in the quasar spectrum; however, we note that most of the group is offset from the quasar line of sight by $\sim$3 $h^{-1}\,$cMpc.  Damping wing attenuation of the group galaxies would therefore be possible if the line-of-sight extent of the island varies across the field.

\subsection{A line-of-sight enhancement of Ly$\alpha$ visbility?}
\label{subsection:line_of_sight}

In contrast to the $z \simeq 6.19$ group, where we see relatively little emission, we find a high rate of rate of \lya\ in two groups at $z \simeq 5.73$ and 5.78.  In these groups, four (six) galaxies out of the six show Ly$\alpha$ emission with at least 5$\sigma$ (3$\sigma$) confidence. Although the galaxies at $z \simeq 5.73$ have somewhat fainter UV magnitude and \Oiii\ luminosities, those at $z \sim 5.78$ fall on the brighter end of our sample. Given their spatial coincidence and similarity in redshift, these groups may form a filament-like structure extending radially from $z \sim 5.71$ to $z \sim 5.78$. Such a large ($\sim$21 $h^{-1}\,$cMpc) ionized filament could produce a region of locally enhanced ionization, facilitating Ly$\alpha$ transmission.

We note that enhanced transmission of Ly$\alpha$ lines in radially extended ionized filaments has been reported elsewhere. \citet{Chen2024} observed 16 LAEs in three groups at $z \sim 7.2$, $z \sim 7.5$, and $z \sim 7.7$. They suggest that these groups may form an extended structure along the line of sight through which \lya\ photons may redshift and escape.
 \cite{Endsley2022b} detected  Ly$\alpha$ emission in nine UV-bright galaxies at $z \sim 6.701$--6.882, suggesting that the LAEs may reside in a  large ($R \sim 44~h^{-1}\,{\rm cMpc}$) ionized bubble.
 As part of their work using JADES, MUSE, and FRESCO sources, \cite{Witstok2024} also find that extended structures may enhance the transmission of Ly$\alpha$ from galaxies at $z \sim 5.9$ and 7.3.  Our results support the idea that group size and geometry may remain significant factors in the detectability even below redshift six.

\subsection{Metal lines and Ly$\alpha$ detection}
\label{subsection:metal-lines}

Finally, we note that our data allow us to take a first look at the potential connection between metal absorbers along the quasar line of sight and \lya\ emission from nearby galaxies.  There is evidence that the number density of low-ionization metal absorbers declines with decreasing redshift from $z \sim 6$ to 5, an indication that the ionization of circum-galactic metals may at least partly trace the ionization of the IGM near the end of reionization \citep{Becker2019,Sebastian2024}.  If so, then one might expect to see a correlation between metal ionization and the transmission of \lya\ from nearby galaxies.

As shown in Figure \ref{fig:3d-figure}, eight metal absorbers are identified at $z > 5.3$ towards J0100+2802. Three of these systems, at $z_{abs} \approx 5.34$, $6.01$, and $6.19$ show significant absorption by C IV and are considered high-ionization systems, while the five at $z_{abs} \approx 5.35$, $5.80$, $5.95$, $6.11$, and $6.14$ are classified are predominantly low-ionization \citep{Chen2017, Becker2019, Cooper2019, Davies2023}. All of the low-ionization systems except the one at $z_{abs} = 5.35$ show O I absorption and likely trace neutral gas \citep{Becker2019,Cooper2019}. 

\citet{Kashino2023} identified potentially associated \Oiii\ emitters  within $\pm 500$ ~$\mathrm{km~s^{-1}}$ and 300 pkpc of all three high-ionization systems, but only one of the low-ionization systems.  We observed two of these (ID 7979 and 6898) associated with the high-ionization systems at redshifts $\simeq$6.01 and 6.19, as well as one (ID 8200) near the low-ionization system at $z \simeq 5.95$. The only one of these we detected in \lya\ was ID 6898, which is associated with a high-ionization system, and even in that case the detection was tentative (3$\sigma$).  Our sample is too small to draw meaningful conclusions between metal absorber properties and galaxy \lya\ transmission.  We do note, however, that the low-ionization systems are predominantly found in low-density regions, where the local ionizing background may be below average. This is consistent with the idea that low-ionization absorbers tend to trace lower ionization regions \citep{Becker2019,Doughty2019}.

\section{Summary and conclusion}
\label{section:summary}

We have used  Keck-LRIS  to measure Ly$\alpha$ emission from a sample of 46 {\it JWST} NIRSpec-identified \Oiii-emitting galaxies spanning $5.34 \le z \le 6.25$.  The galaxies are drawn from the EIGER catalog in the field of the $z = 6.33$ quasar J0100+2802 \citep{Kashino2023}. Our primary goal has been to investigate the relationship between environment and Ly$\alpha$ visibility. 




In total, we detect 10 galaxies in Ly$\alpha$ emission with at least 5$\sigma$ confidence. While our overall detection rate is consistent with literature results, we find significant variation between groups.  Notably, we find evidence that a large group at $z \simeq 6.19$ may exhibit significantly (factor of $\sim$3) less transmission than the remainder of the sample.  The evidence comes primarily from the ratios of \lya\ and \Oiii\ luminosities, although there is tentative evidence that the \lya\ equivalent width distribution may also be different.  The apparent deficit of \lya\ in this group is significant given its size and the fact that it falls within the redshift of the quasar's proximity zone, although most of the group members are significantly offset from the quasar line of sight.  We speculate that a foreground neutral island may be scattering \lya\ photons from the galaxies in this group.   
In contrast, we find a high rate of \lya\ detections in a pair of closely separated groups at $z \simeq 5.73$ and $z \simeq 5.78$.  In this case, the geometric arrangement of an extended structure may be facilitating \lya\  transmission.

While it is difficult to draw firm conclusions from this limited sample, our study suggests that environment may play a significant role in the visibility of \lya\ emission even as late as $z \sim 6$.  If confirmed in larger studies, this could further increase the ability of galaxy \lya\ emission to serve as a probe of the end stages of reionization.

\section*{Acknowledgements}

We would like to thank Dr. Jorryt Matthee and the EIGER team for generously sharing the \Oiii, UV and  JWST photometric data for this sample. The authors also would like to thank Dr. Brian Siana for helpful comments.  SH and GB were supported by NASA through a grant from the Space Telescope Science Institute grant for {\it JWST} program GO-4092.

Some of data  presented herein were obtained at the W. M. Keck Observatory, which is
operated as a scientific partnership among the California Institute of Technology,
the University of California and the National Aeronautics and Space Administration. The Observatory was made possible by the generous financial support
of the W. M. Keck Foundation. The authors wish to recognize and acknowledge
the very significant cultural role and reverence that the summit of Maunakea has
always had within the indigenous Hawaiian community.

The authors would like to thank our Keck Observatory
Support Astronomers, 
Percy Gomez and Chien-Hsiu Lee,
for their assistance during our observing runs. Special
thanks to our Observing Assistant, Arina Rostopchina, for her work.

This work is based in part on observations made with the NASA/ESA/CSA James Webb Space Telescope. These observations are associated with program \#1243.

This work made use of the following softwares: numpy (\citealt{Harris2020}), matplotlib (\citealt{Hunter2007}), SpectRes (\citealt{Carnall2017}), scipy (\citealt{Virtanen2020}),
astropy, a community-developed core Python package for Astronomy (\citealt{AstropyCollaboration2013, AstropyCollaboration2020}).

\section*{Data Availability}

The {\it JWST} data used here are available on the Mikulski Archive for Space Telescopes (\url{https://mast.stsci.edu/}). 
Other data will be shared upon reasonable request to the corresponding author.


\bibliographystyle{mnras}
\bibliography{lae_ref} 



\appendix

\section{Some extra material}

\subsection{1D and 2D  spectra of our sample }
\label{subsec:fullsamplespectra}
\begin{figure*}
    \centering
    \includegraphics[width=0.331\textwidth]{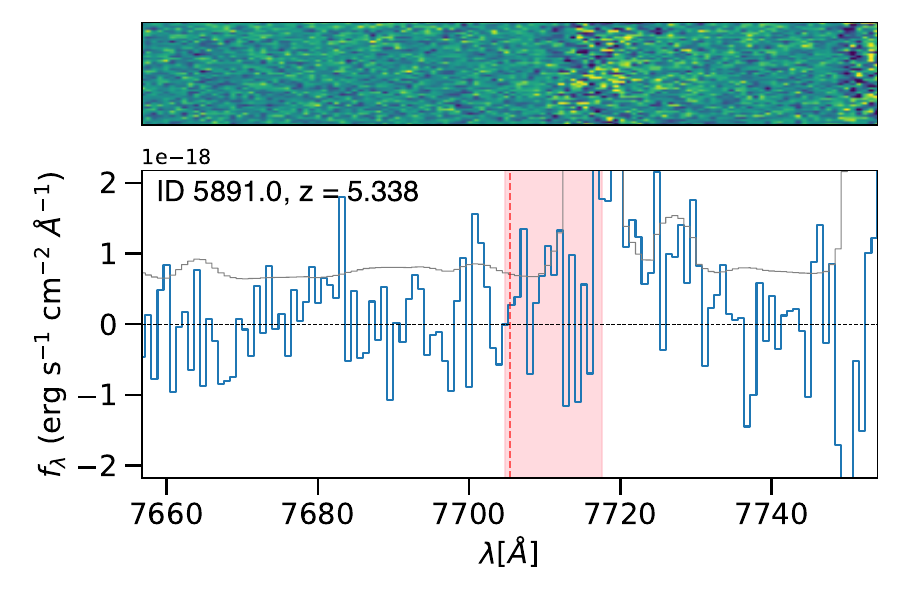}
    \hfill
    \includegraphics[width=0.331\textwidth]{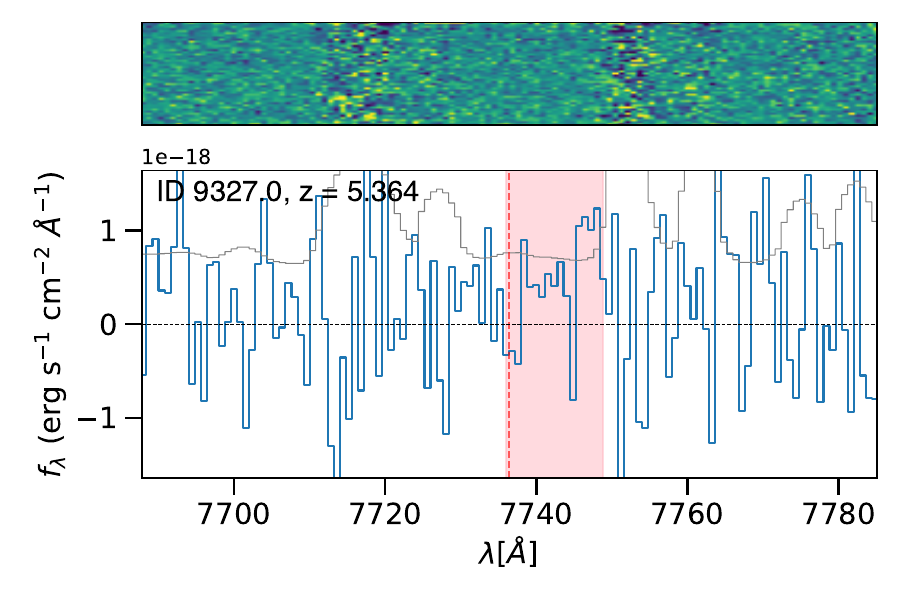}
    \hfill
    \includegraphics[width=0.331\textwidth]{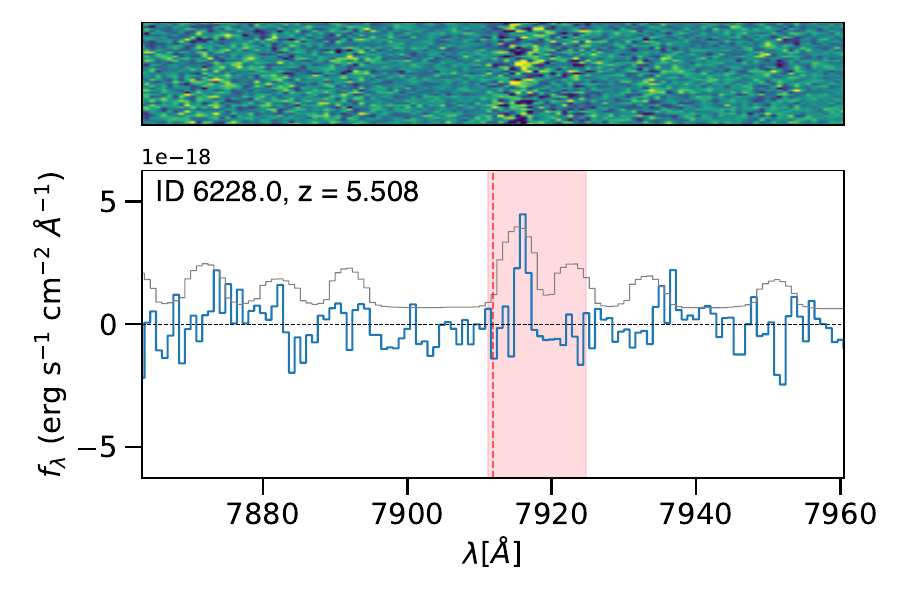}

    \vspace{0.3cm}

    \includegraphics[width=0.331\textwidth]{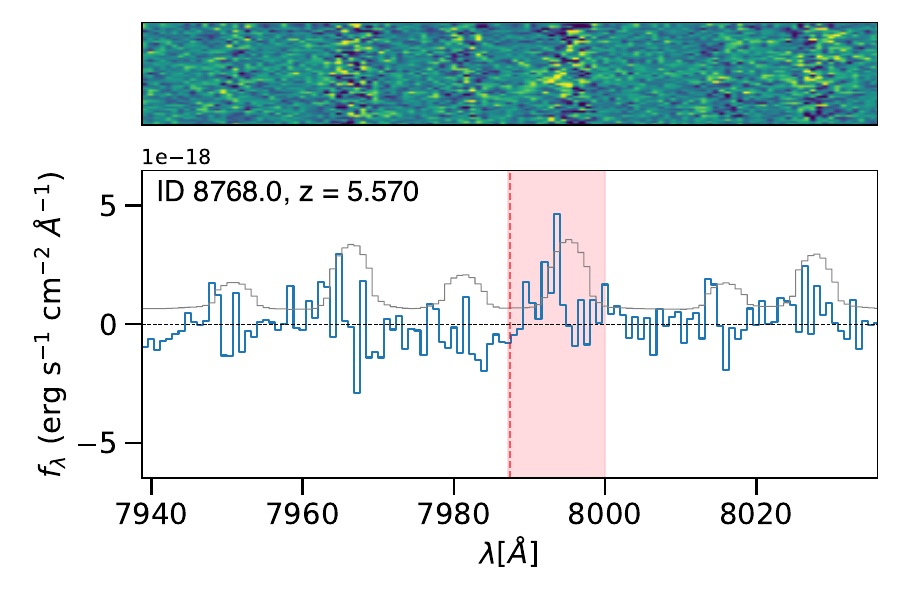}
    \hfill
    \includegraphics[width=0.331\textwidth]{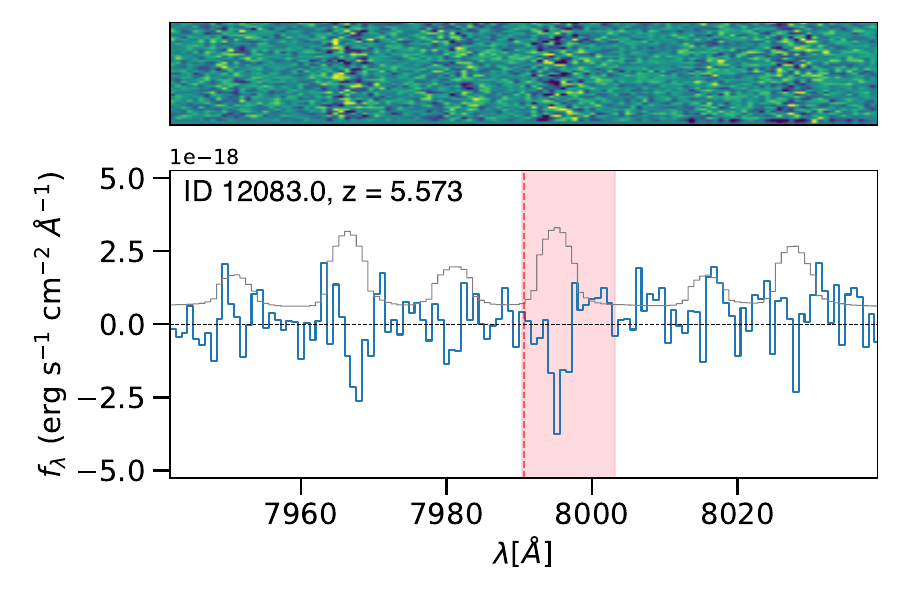}
    \hfill
    \includegraphics[width=0.331\textwidth]{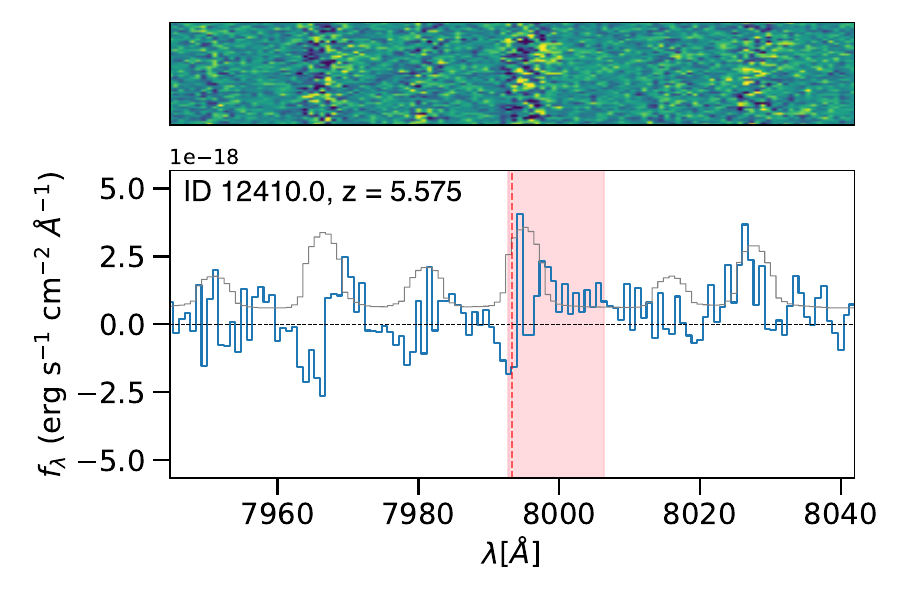}

    \vspace{0.3cm}

    \includegraphics[width=0.331\textwidth]{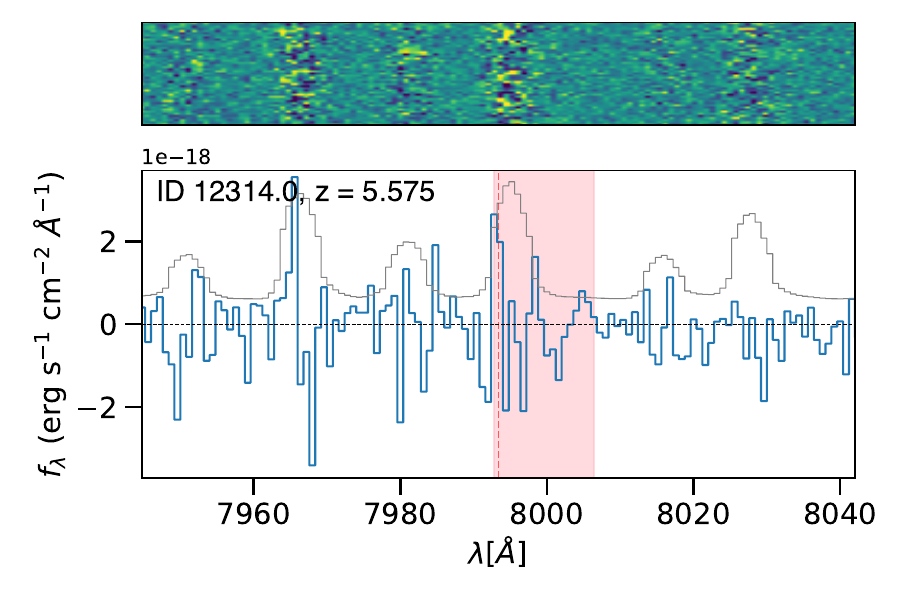}
    \hfill
    \includegraphics[width=0.331\textwidth]{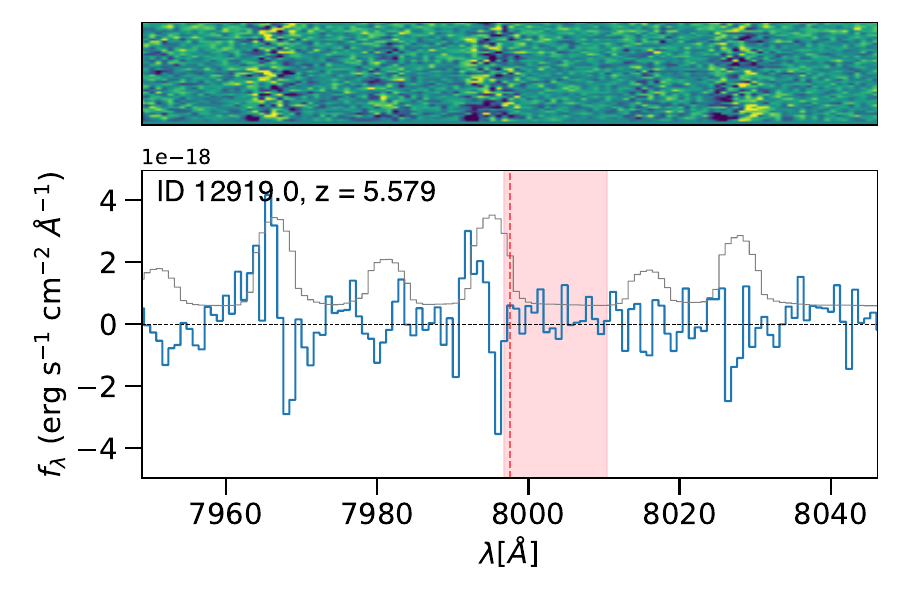}
    \hfill
    \includegraphics[width=0.331\textwidth]{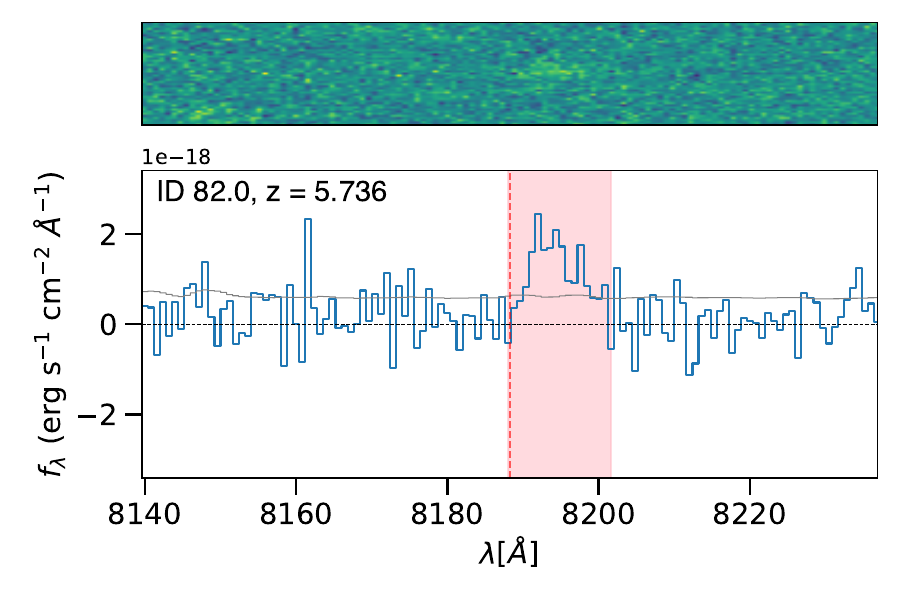}

    \vspace{0.3cm}

    \includegraphics[width=0.331\textwidth]{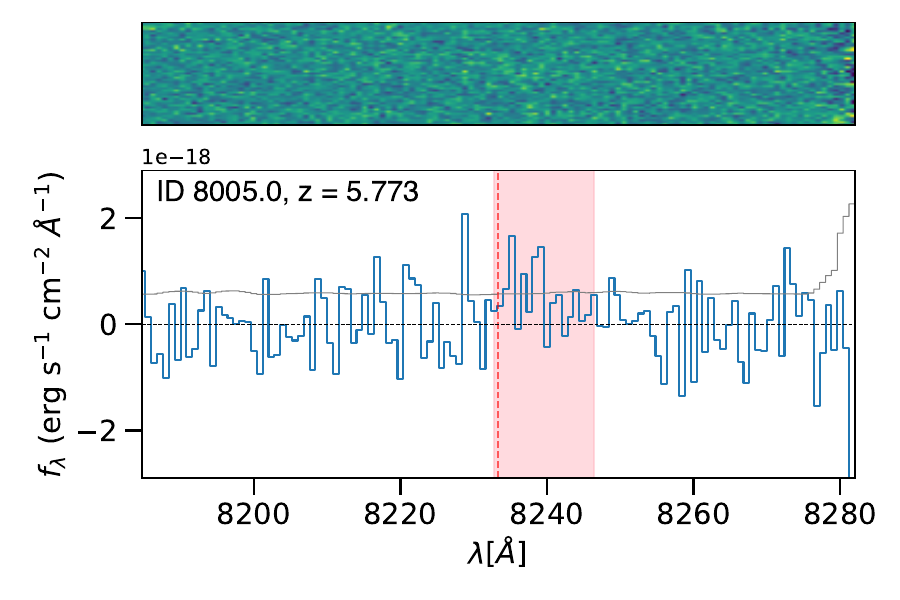}
    \hfill
    \includegraphics[width=0.331\textwidth]{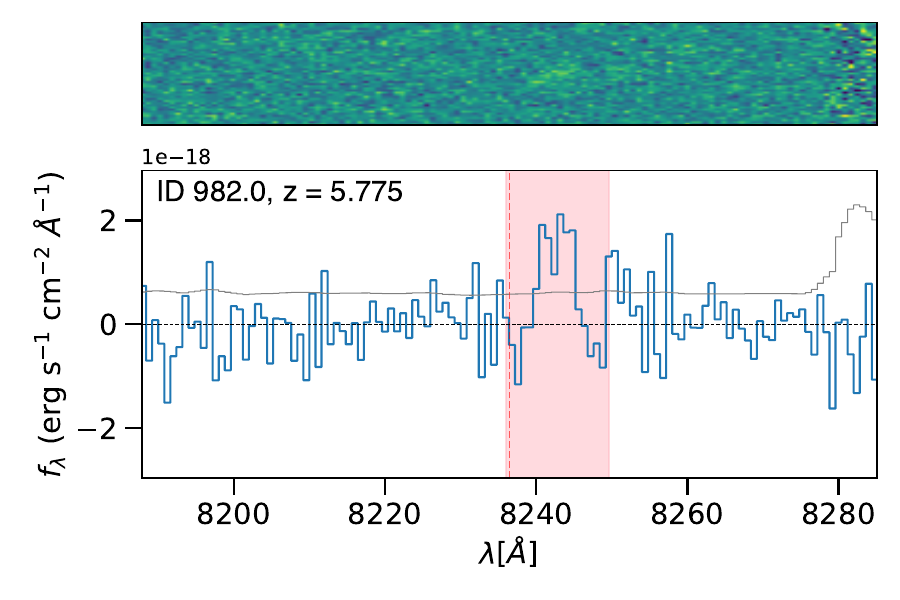}
    \hfill
    \includegraphics[width=0.331\textwidth]{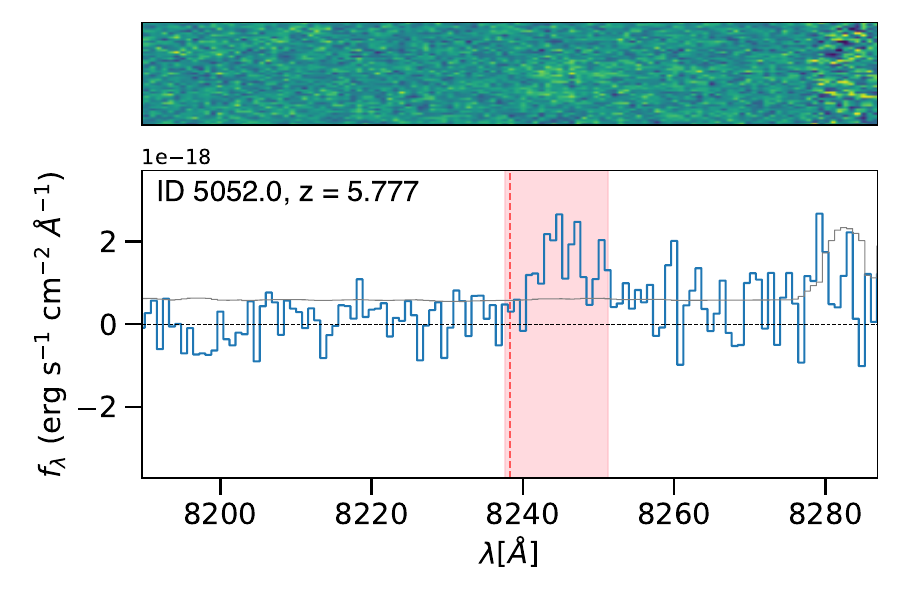}

    \vspace{0.3cm}

    \includegraphics[width=0.331\textwidth]{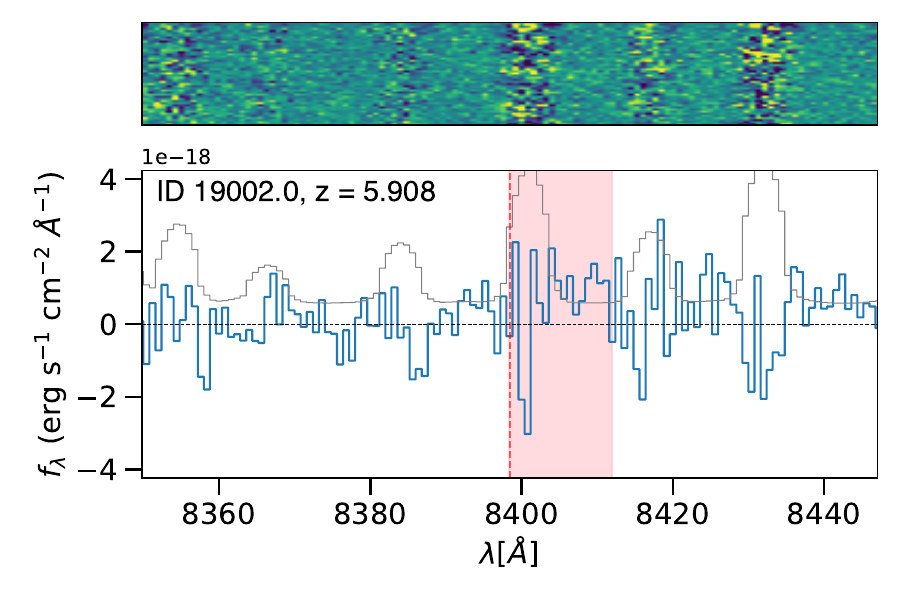}
    \hfill
    \includegraphics[width=0.331\textwidth]{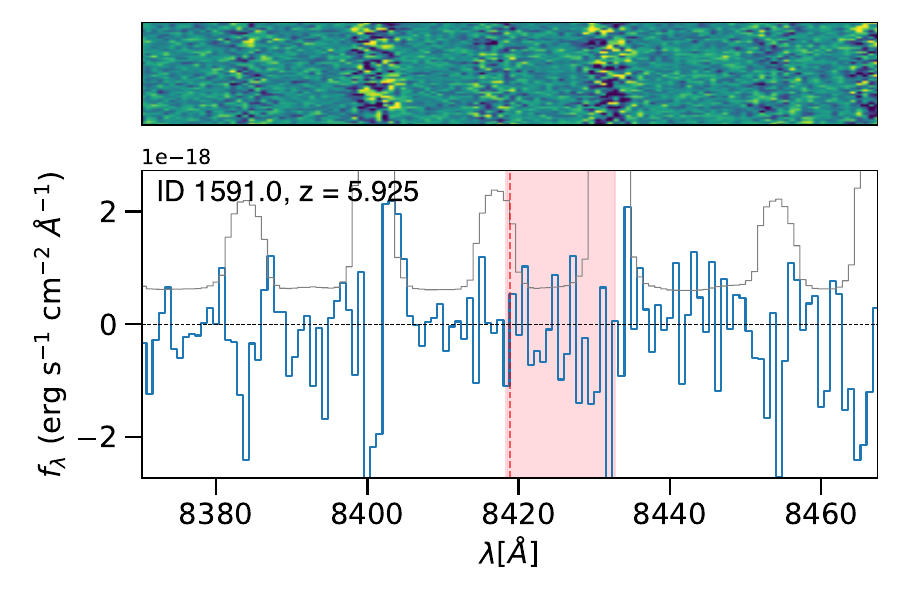}
    \hfill
    \includegraphics[width=0.331\textwidth]{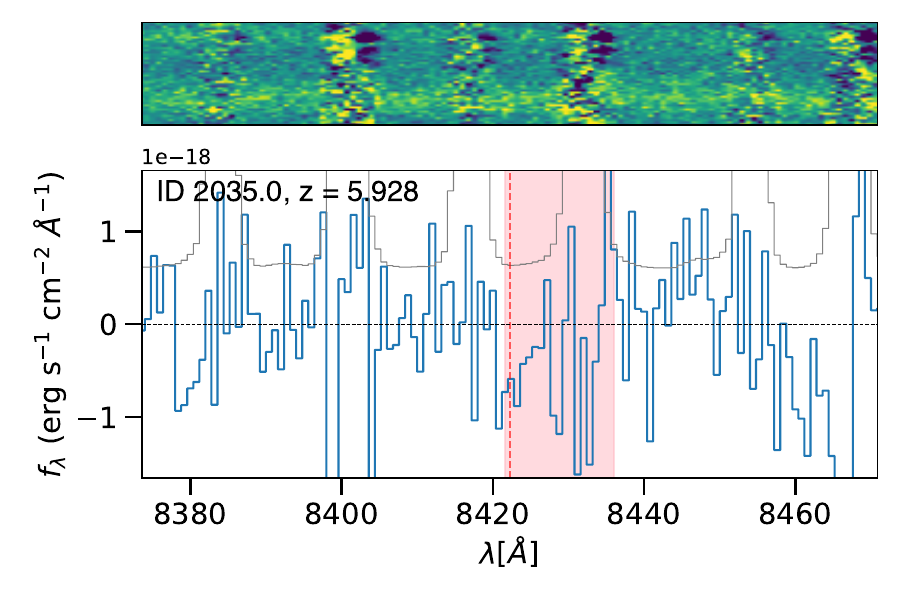}

    \caption{As in Figure \ref{fig:representativesamples} but for the remainder of the sample. }
    \label{fig:nondetection1}
\end{figure*}

\begin{figure*}
    \centering
    \includegraphics[width=0.331\textwidth]{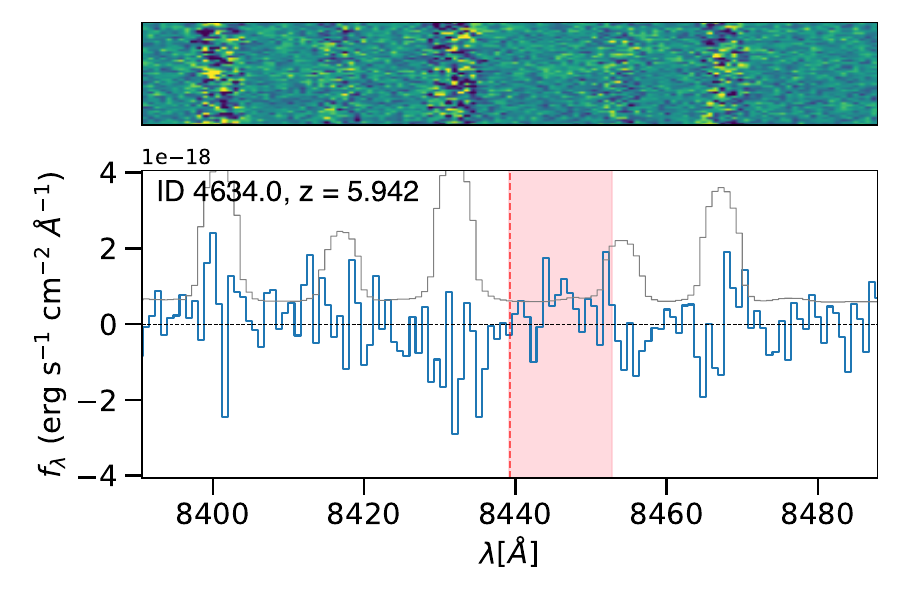}
    \hfill
    \includegraphics[width=0.331\textwidth]{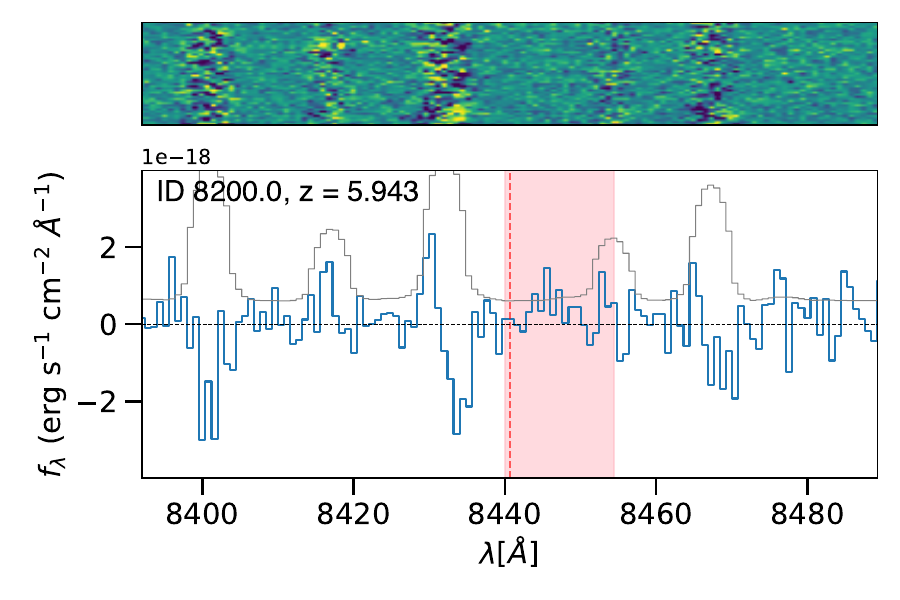}
    \hfill
    \includegraphics[width=0.331\textwidth]{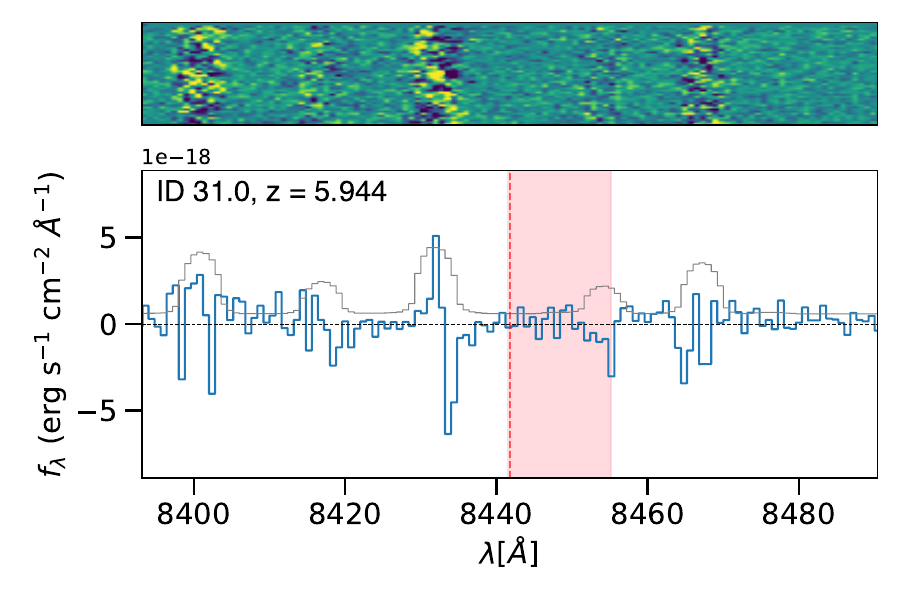}

    \vspace{0.3cm}

    \includegraphics[width=0.331\textwidth]{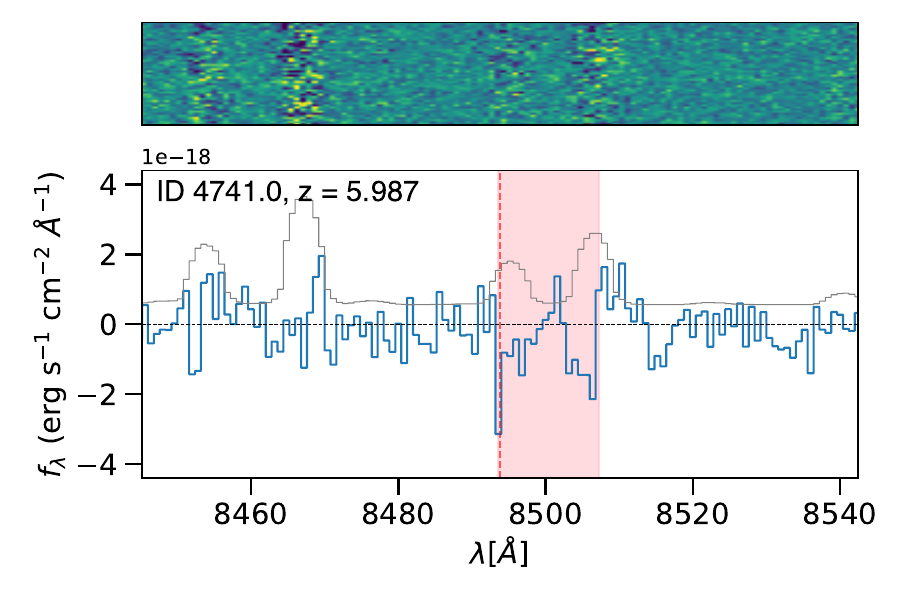}
    \hfill
    \includegraphics[width=0.331\textwidth]{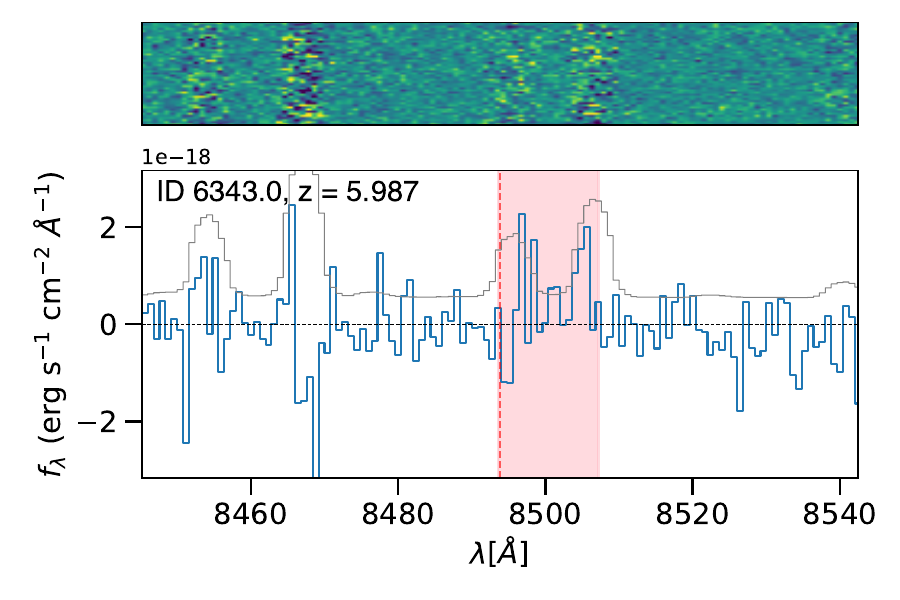}
    \hfill
    \includegraphics[width=0.331\textwidth]{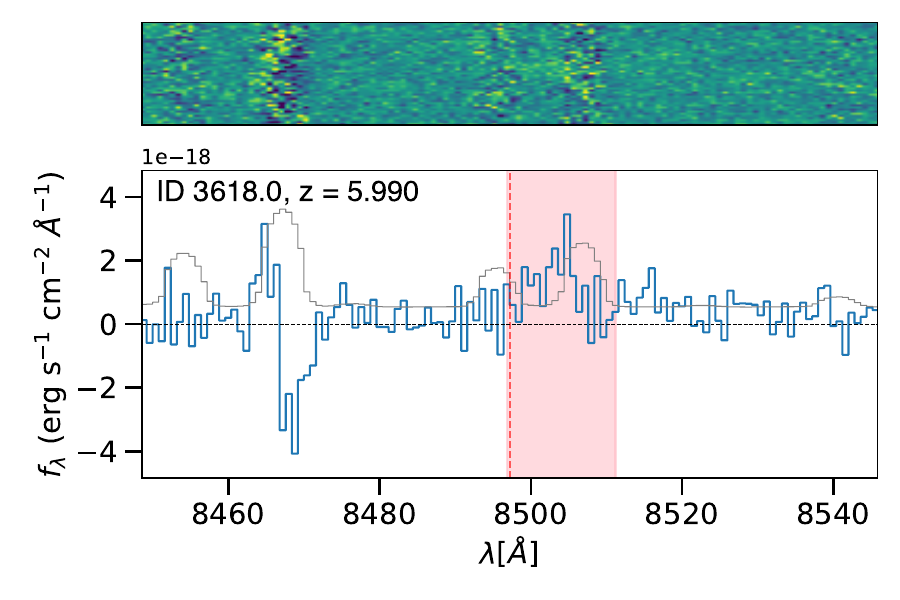}

    \vspace{0.3cm}

    \includegraphics[width=0.331\textwidth]{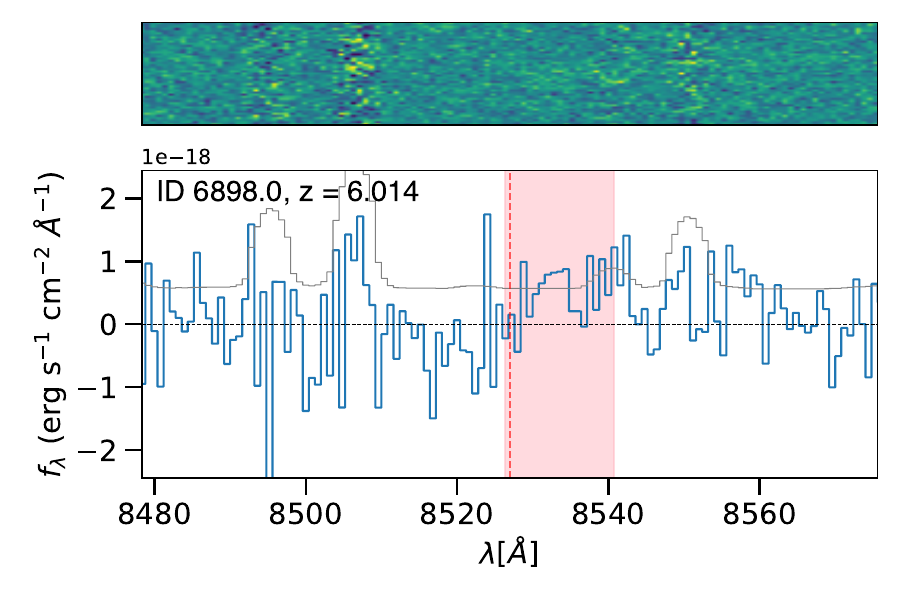}
    \hfill
    \includegraphics[width=0.331\textwidth]{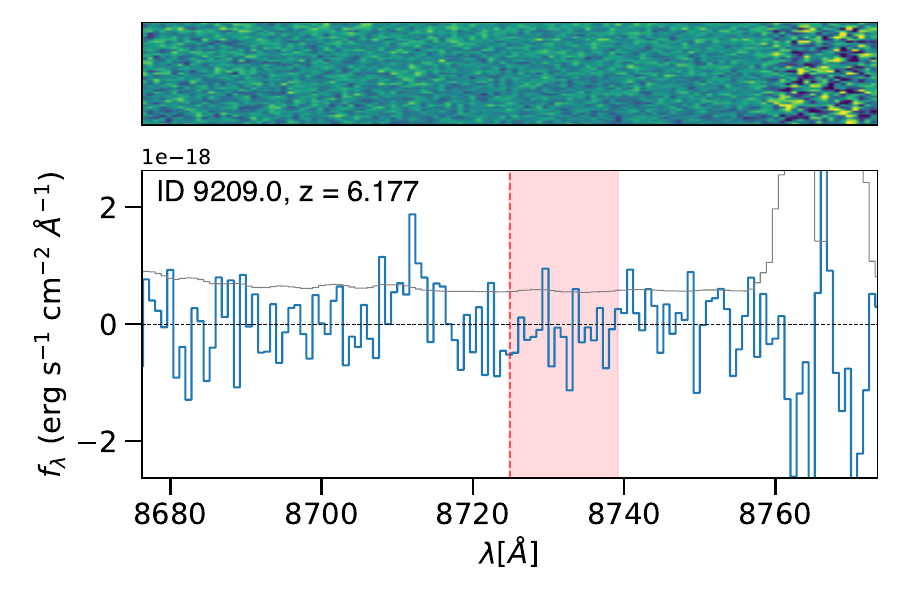}
    \hfill
    \includegraphics[width=0.331\textwidth]{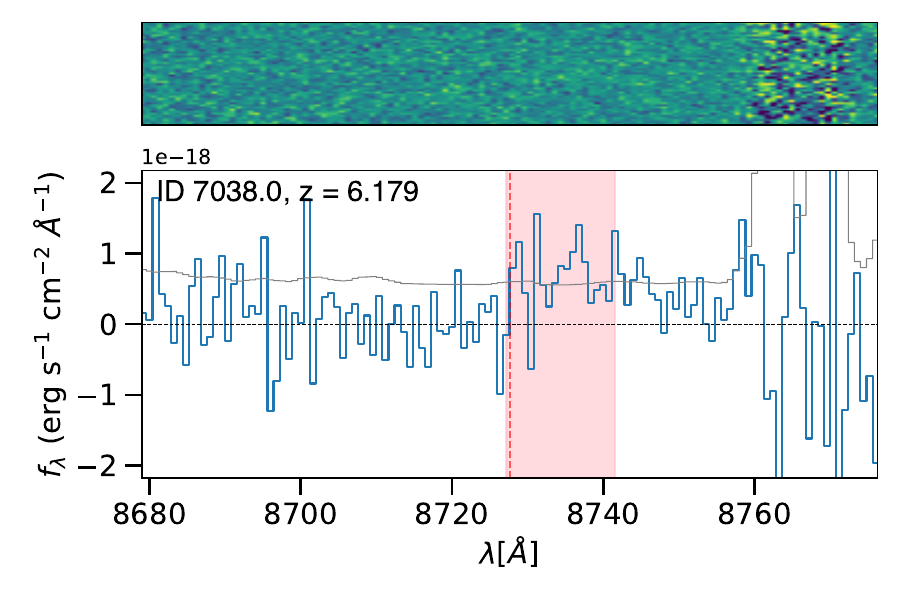}

    \vspace{0.3cm}

    \includegraphics[width=0.331\textwidth]{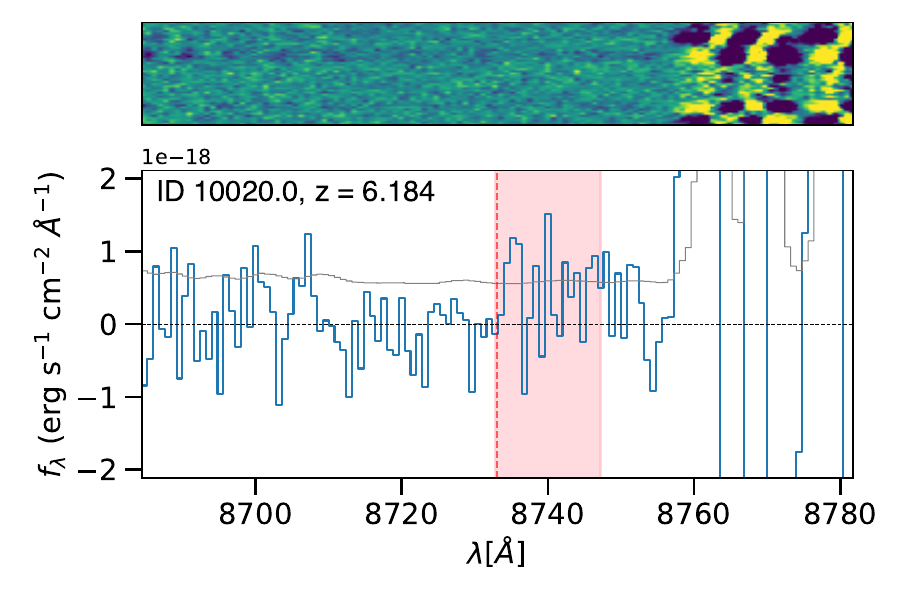}
    \hfill
    \includegraphics[width=0.331\textwidth]{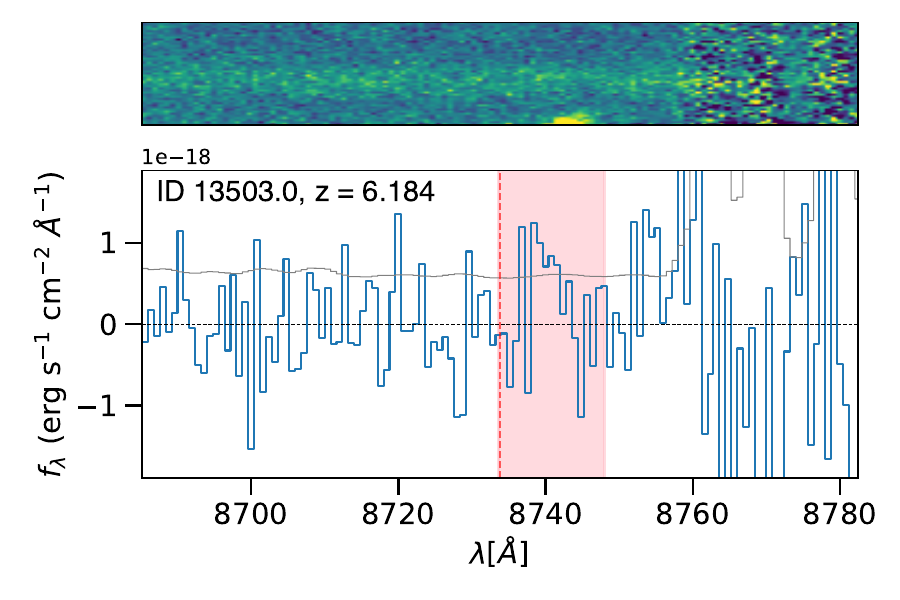}
    \hfill
    \includegraphics[width=0.331\textwidth]{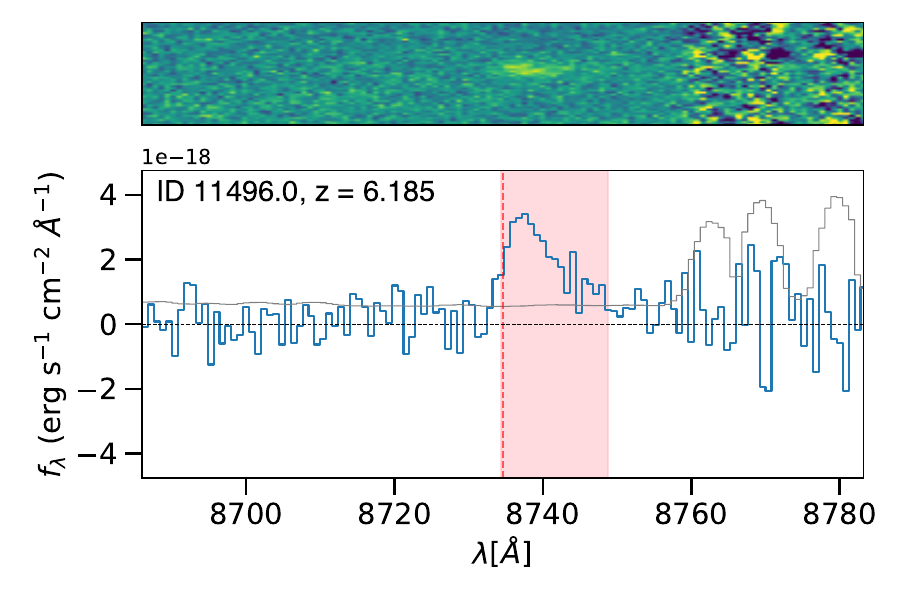}

    \vspace{0.3cm}

    \includegraphics[width=0.331\textwidth]{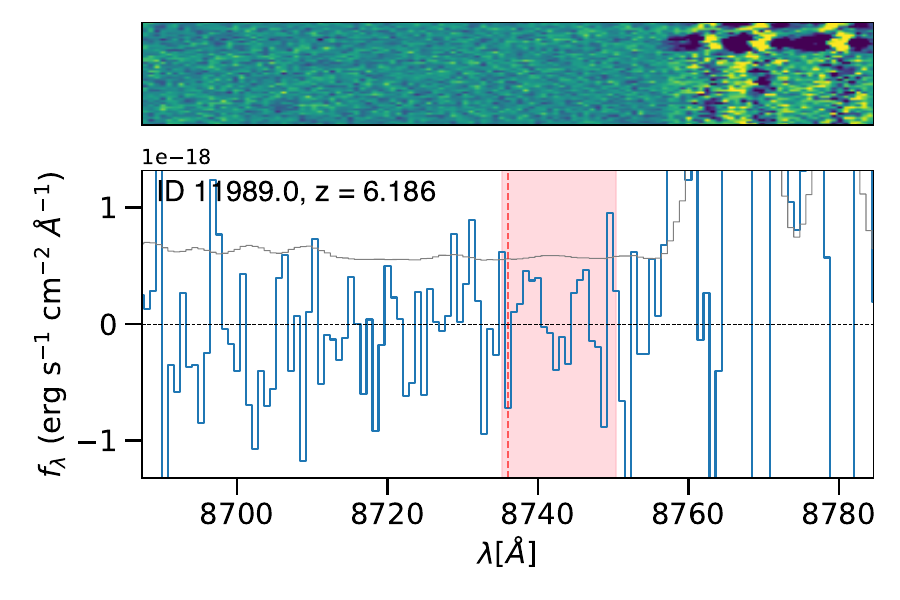}
    \hfill
    \includegraphics[width=0.331\textwidth]{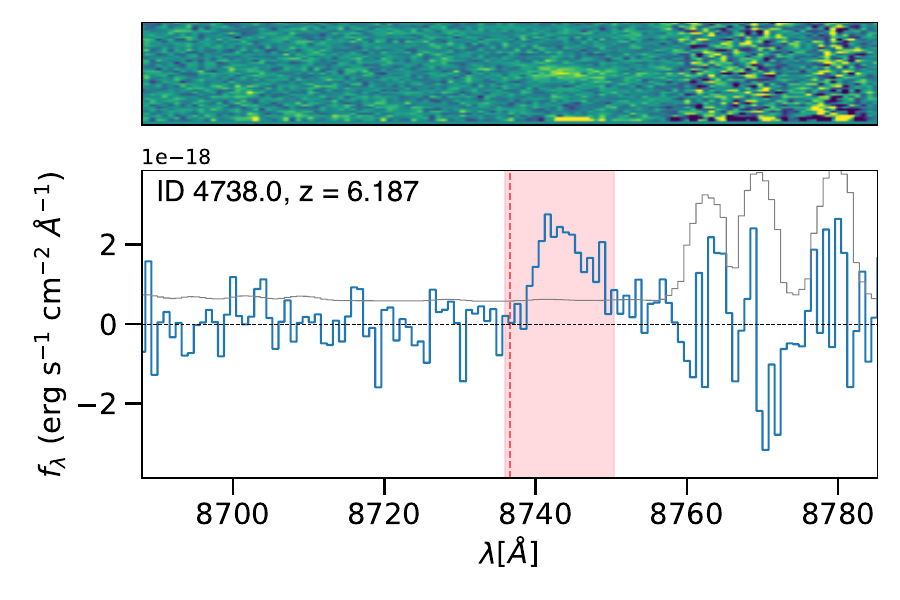}
    \hfill
    \includegraphics[width=0.331\textwidth]{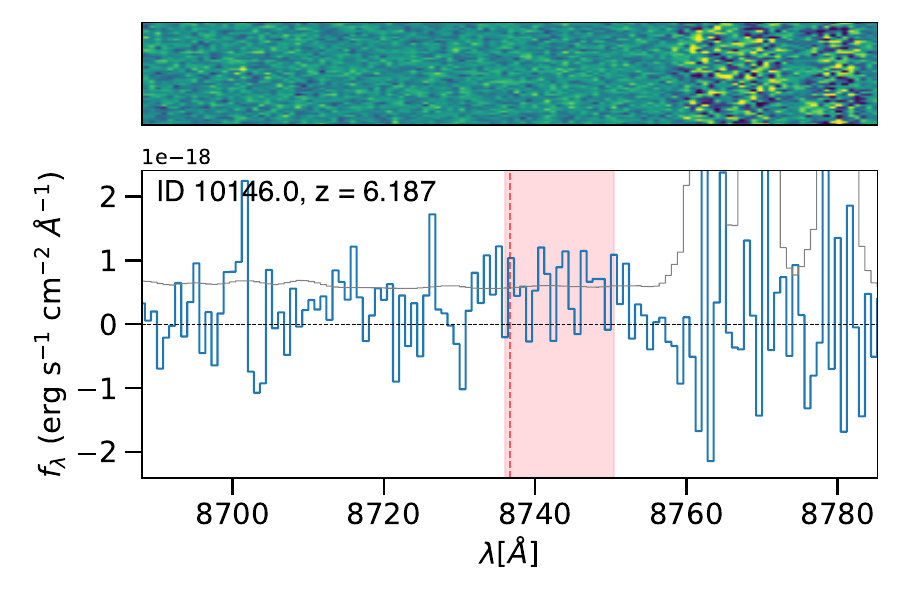}

    \caption{ Figure \ref{fig:nondetection1} continued. In some cases (e.g., object IDs 10020 and 11989), strong and irregular skyline residuals are caused by small irregularities in the slit. }
    \label{fig:nondetection2}
\end{figure*}

\begin{figure*}
    \centering
    \includegraphics[width=0.331\textwidth]{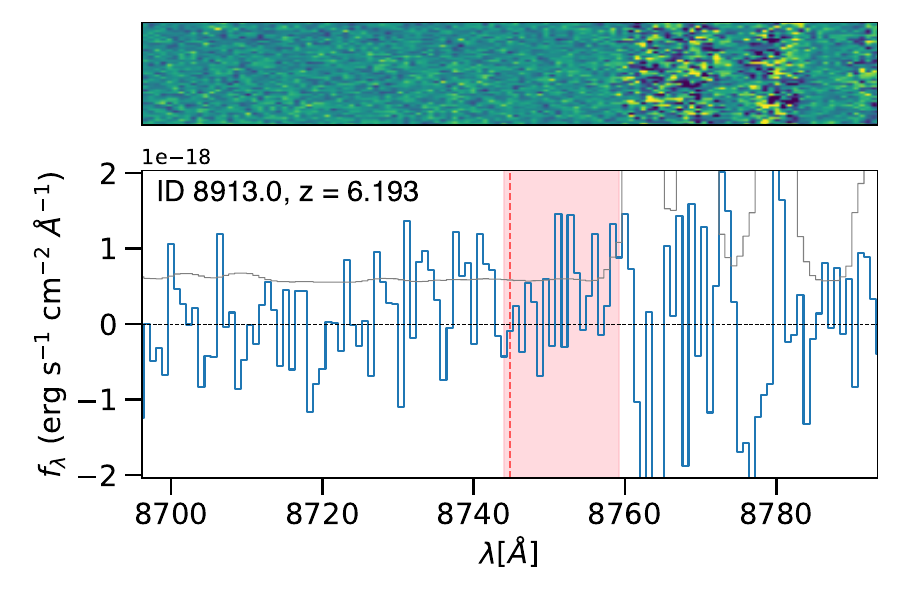}
    \hfill
    \includegraphics[width=0.331\textwidth]{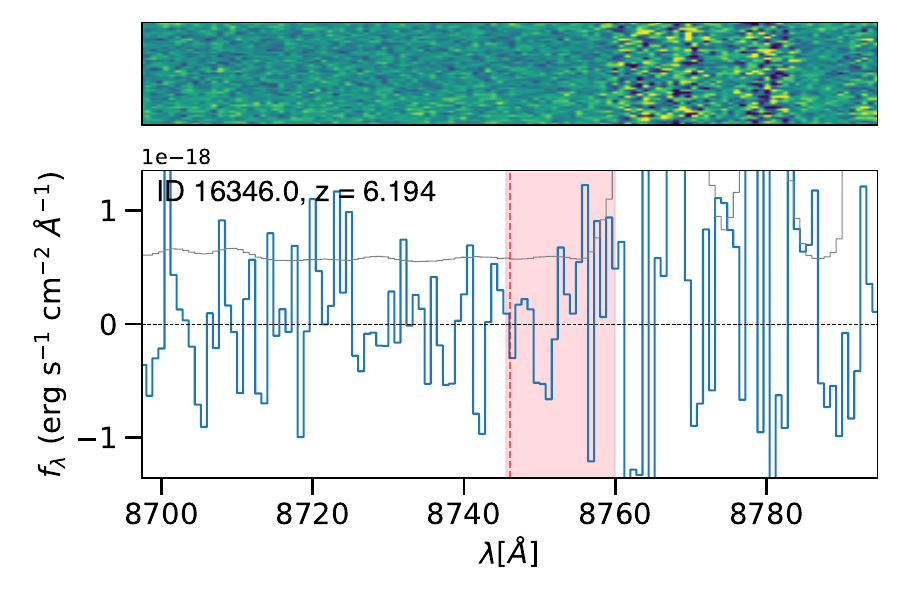}
    \hfill
    \includegraphics[width=0.331\textwidth]{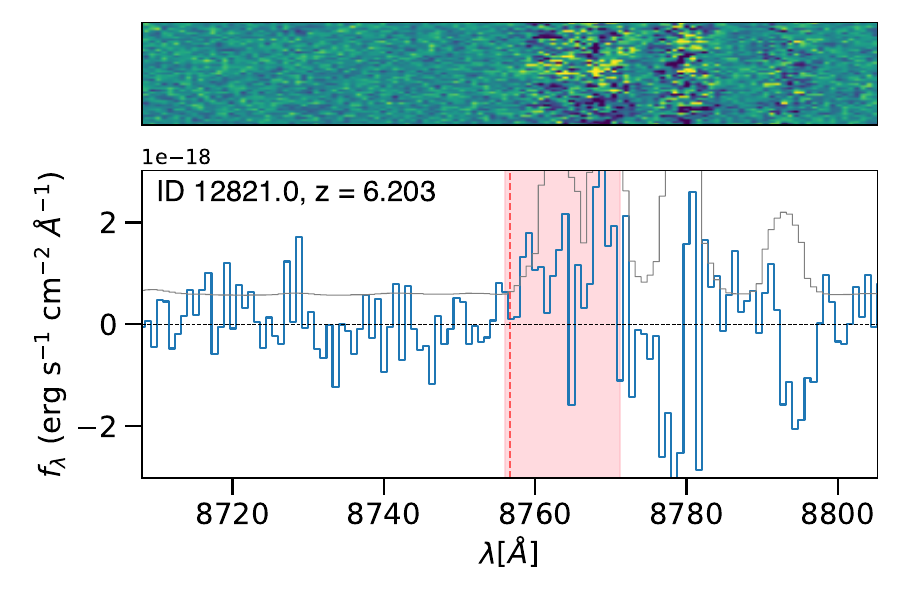}

    \vspace{0.3cm}

    \includegraphics[width=0.331\textwidth]{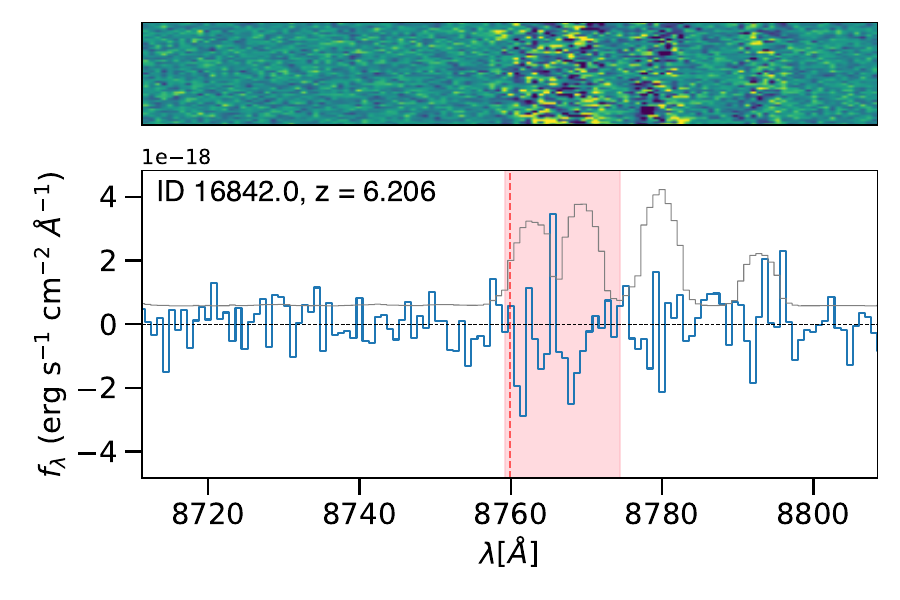}
    \hfill
    \includegraphics[width=0.331\textwidth]{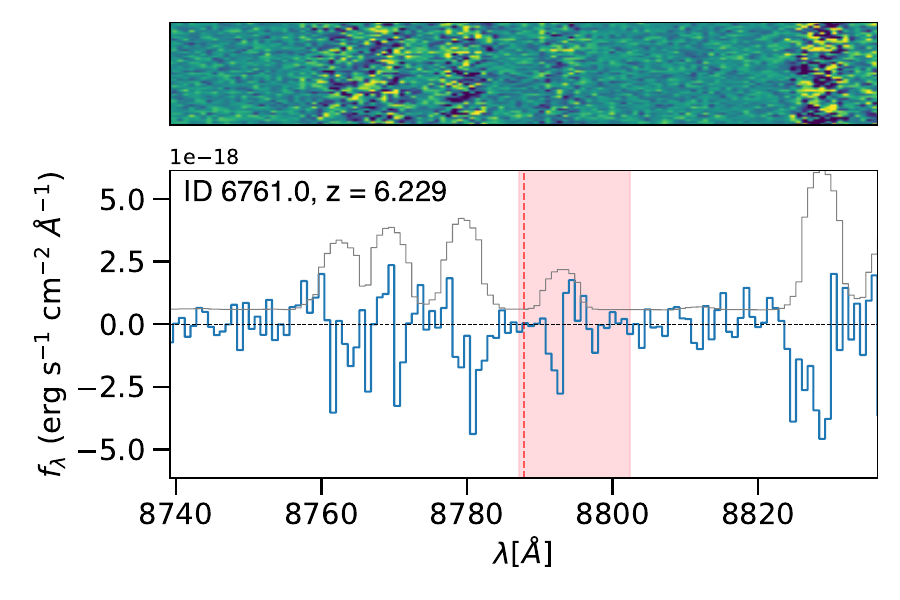}
    \hfill
    \includegraphics[width=0.331\textwidth]{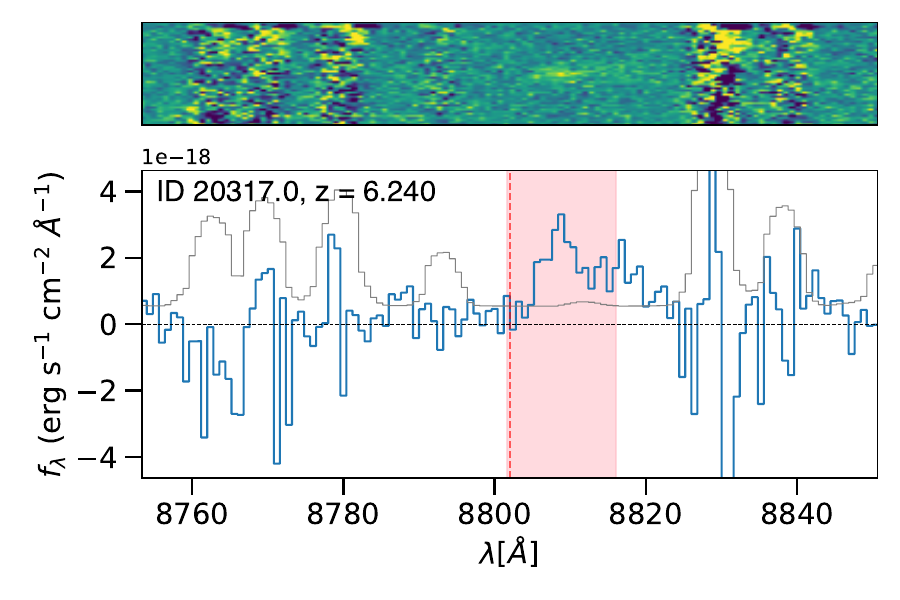}

    \vspace{0.3cm}
    \includegraphics[width=0.331\textwidth]{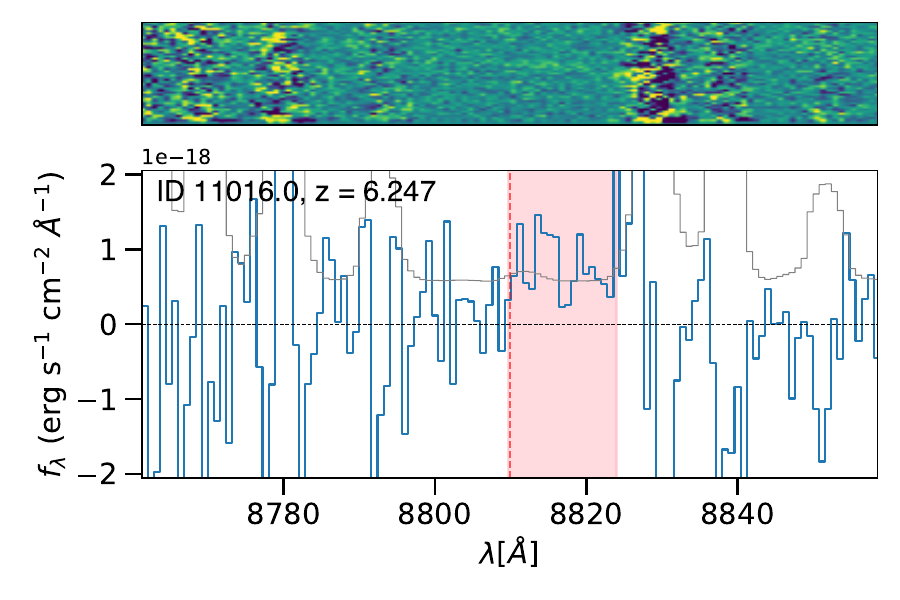}
    \hfill

    \caption{ Figure \ref{fig:nondetection1} continued }
    \label{fig:nondetection3}
\end{figure*}

In Figures \ref{fig:nondetection1}, \ref{fig:nondetection2}  and \ref{fig:nondetection3}, we show the 1D and 2D  spectra of the remaining of our  sample (37 galaxies which are not shown in the main text).

\bsp	
\label{lastpage}
\end{document}